\documentclass[12pt,preprint]{aastex}

\usepackage{ulem}
\usepackage{color}

\newcommand{\tma}[2]{\if0#1\fi#2}					

\newcommand{\tmb}[2]{\if0#1\fi#2}					

\shorttitle{Dense clumps in GMCs in LMC}
\shortauthors{Minamidani et al.}

\begin{document}


\title{
Dense Clumps in Giant Molecular Clouds in the Large Magellanic Cloud:\\
Density and Temperature Derived from $^{13}$CO($J=3-2$) Observations
}


\author{
Tetsuhiro Minamidani\altaffilmark{1,2}, 
Takanori Tanaka\altaffilmark{3}, 
Yoji Mizuno\altaffilmark{3}, 
Norikazu Mizuno\altaffilmark{4}, 
Akiko Kawamura\altaffilmark{3}, 
Toshikazu Onishi\altaffilmark{5}, 
Tetsuo Hasegawa\altaffilmark{4}, 
Ken'ichi Tatematsu\altaffilmark{4}, 
Tatsuya Takekoshi\altaffilmark{2}, 
Kazuo Sorai\altaffilmark{1,2}, 
Nayuta Moribe\altaffilmark{3}, 
Kazufumi Torii\altaffilmark{3}, 
Takeshi Sakai\altaffilmark{6}, 
Kazuyuki Muraoka\altaffilmark{5}, 
Kunihiko Tanaka\altaffilmark{7}, 
Hajime Ezawa\altaffilmark{4}, 
Kotaro Kohno\altaffilmark{6}, 
Sungeun Kim\altaffilmark{8},
M\'{o}nica Rubio\altaffilmark{9},
\and
Yasuo Fukui\altaffilmark{3}
}

\email{tetsu@astro1.sci.hokudai.ac.jp}


\altaffiltext{1}{Department of Physics, Faculty of Science, Hokkaido University, N10W8, Kita-ku, Sapporo 060-0810, Japan}
\altaffiltext{2}{Department of Cosmosciences, Graduate School of Science, Hokkaido University, N10W8, Kita-ku, Sapporo 060-0810, Japan}
\altaffiltext{3}{Department of Astrophysics, Nagoya University, Furo-cho, Chikusa-ku, Nagoya 464-8602, Japan}
\altaffiltext{4}{National Astronomical Observatory of Japan, Mitaka, Tokyo 181-8588, Japan}
\altaffiltext{5}{Department of Physical Science, Osaka Prefecture University, Gakuen 1-1, Sakai, Osaka 599-8531, Japan}
\altaffiltext{6}{Institute of Astronomy, The University of Tokyo, 2-21-1 Osawa, Mitaka, Tokyo 181-0015, Japan}
\altaffiltext{7}{Institute of Science and Technology, Keio University, 4-14-1 Hiyoshi, Yokohama, Kanagawa 223-8522, Japan}
\altaffiltext{8}{Astronomy \& Space Science Department, Sejong University, 98 Kwangjin-gu, Kunja-dong, Seoul, 143-747, Korea}
\altaffiltext{9}{Departament de Astronom\'ia, Universidad de Chile, Casilla 36-D, Santiago, Chile}


\begin{abstract}
In order to precisely determine temperature and density of molecular gas in the Large Magellanic Cloud, we made observations of optically thin $^{13}$CO($J=3-2$) transition by using the ASTE 10m telescope toward 9 peaks \tma{of }{where} $^{12}$CO($J=3-2$) clumps \tma{}{were} previously detected with the same telescope. The molecular clumps include those in giant molecular cloud (GMC) Types I (with no signs of massive star formation), II (with HII regions only), and III (with HII regions and young star clusters). We detected $^{13}$CO($J=3-2$) emission toward all the peaks and found that their intensities are 3 -- 12 times lower than those of $^{12}$CO($J=3-2$). We \tma{derived }{determined} the intensity ratios of $^{12}$CO($J=3-2$) to $^{13}$CO($J=3-2$), $R^{12/13}_{3-2}$, and $^{13}$CO($J=3-2$) to $^{13}$CO($J=1-0$), $R^{13}_{3-2/1-0}$, at 45$\arcsec$ resolution\tma{,}{.} \tma{and these }{These} ratios \tma{are }{were} used for radiative transfer calculations in order to estimate temperature and density of the clumps. \tma{The clumps range from 15K to 200K in kinetic temperature, and from 8$\times 10^2$ to 7$\times 10^3$ cm$^{-3}$ in density.}{The parameters of these clumps range kinetic temperature $T\mathrm{_{kin}}$ = 15 -- 200 K, and molecular hydrogen gas density $n(\mathrm{H_2})$ = 8$\times 10^2$ -- 7$\times 10^3$ cm$^{-3}$.} We confirmed that the higher density clumps show higher kinetic temperature and that the lower density clumps lower kinetic temperature at a better accuracy than in the previous work. The kinetic temperature and density increase generally from a Type I GMC to a Type III GMC. We interpret that this difference reflects an evolutionary trend of star formation in molecular clumps. The $R^{13}_{3-2/1-0}$ and kinetic temperature of the clumps are well correlated with H$\alpha$ flux, suggesting that the heating of molecular gas \tma{whose density is }{$n(\mathrm{H_2})$ =} $10^3$ -- $10^4$ cm$^{-3}$ can be explained by stellar FUV photons.
\end{abstract}





\section{Introduction}

Star formation is of fundamental importance in understanding the evolution of galaxies. Stars are formed in dense clumps of giant molecular clouds (GMCs)\tma{}{,} and kinetic energy and heavy elements are ejected from stars back into the interstellar medium (ISM) through stellar winds and supernova explosions. This cycle enriches metal abundance in the ISM and drives the evolution of galaxies both physically and chemically. It is therefore important to better understand the evolution of GMCs, the principle sites of the star formation in galaxies.

The Large Magellanic Cloud (LMC) is the most suitable galaxy to study star formation and natal GMCs because of its ideal location\tma{ to us}{}. The LMC offers a unique opportunity to achieve the highest \tma{resolutions }{resolution} due to its proximity, 50kpc \citep{Keller2006, Feast1999}, and \tma{its}{} nearly face-on position with an inclination angle of $\sim$ 35\degr  \citep{vanderMarel2001}\tma{}{. This also} provides an advantage of less contamination along the line of sight. The LMC shows active on-going star formation and many massive young clusters are being formed \citep{Hodge1961}. In the LMC, the metallicity is a factor of $\sim$ 3 -- 4 lower \citep{Dufour1984, Rolleston2002} and the gas-to-dust ratio is a factor of $\sim$ 4 higher \citep{Koornneef1984} than \tma{those of }{values for} the solar neighborhood. The far-ultraviolet (FUV) radiation field is more intense in the LMC than in the Milky Way \citep{Israel1986} and the visual extinction is a few times lower than in the Milky Way. These \tma{characterize }{influence} the physical properties of GMCs, and may affect the initial condition of star and cluster formation.

The first spatially resolved surveys of GMCs in the whole LMC were made by the NANTEN 4m telescope in the $^{12}$CO($J=1-0$) transition \citep{Fukui1999, Fukui2001, Fukui2008, MizunoN2001}. These surveys revealed the distribution of the GMCs within a single galaxy at a $\sim$ 40 pc resolution. \citet{Fukui2008} derived the physical properties, such as size, line width, and virial mass, of the GMCs, and found that the $^{12}$CO($J=1-0$) luminosity and virial mass of the clouds show a good correlation. Assuming \tma{}{that} the clouds are in virial equilibrium, \citet{Fukui2008} derived the $X_{\mathrm{co}}$ factor, conversion factor of the $^{12}$CO($J=1-0$) intensity to total molecular column density. The derived $X_{\mathrm{co}}$ factor was similar to that of other Local Group galaxies, such as Small Magellanic Cloud, M31, M33, IC10, and Milky Way, and the GMC mass distribution, dN/dM, was also similar to that of M31, M33, and IC10, suggesting that GMCs in the Local Group galaxies have similar properties \citep{Fukui2010, Blitz2007}. Comparisons between the GMCs and signs of star formation such as HII regions and young stellar clusters were used to classify GMCs into three types in terms of star formation activities \citep{Fukui1999, Kawamura2009}; Type I shows no signs of massive star formation, Type II associated with only small HII regions, and Type III associated with both HII regions and young stellar clusters, and these Types are interpreted as an evolutionary sequence \citep{Kawamura2009}. Comparative studies of \tma{}{the} $^{12}$CO($J=1-0$) to HI \tma{}{ratio} \tma{is }{are} also a key to understand the evolution of GMCs and the GMC formation via conversion of HI into H$_2$. \citet{Wong2009} made \tma{a}{} two dimensional, i.e., spatial, comparison\tma{}{s} of the second NANTEN $^{12}$CO($J=1-0$) survey \citep{Fukui2008} with \tma{ATCA+Parkes HI data sets }{HI data set combined ATCA and Parkes Telescope surveys }\citep{Kim2003}. They found that significant HI column \tma{density }{densities} ($>$ 10$^{21}$ cm$^{-2}$) and peak brightness \tma{temperature }{temperatures} ($>$ 20K) are necessary but not sufficient conditions for CO detection. \citet{Fukui2009} compared \tma{the same data sets in three dimensions }{the three dimensional data cubes}, including a velocity axis in addition to the two spatial axes. They found that GMCs are associated with HI envelopes on scales of $\sim$ 50 -- 100pc, and the HI envelopes may be in dynamical equilibrium or may be accreting onto GMCs to increase the mass via HI -- H$_2$ conversion. \tma{A subsequent }{Subsequently a} higher resolution (45$\arcsec$  $\sim$ 10 pc at 50kpc) survey of $^{12}$CO($J=1-0$) molecular cloud with the Mopra 22m telescope, the Magellanic Mopra Assessment (MAGMA), revealed \tma{}{a} more detailed distribution of the $^{12}$CO($J=1-0$) emission in the individual GMCs \citep{Hughes2010}. \citet{Hughes2010} presented that the physical properties of star-forming GMCs are very similar to the properties of GMCs without signs of massive star formation. 

The physical properties such as kinetic temperature and density of the molecular gas in the LMC have been investigated in the higher-$J$ transitions ($J=2-1$, $J=3-2$, $J=4-3$, $J=7-6$) of CO \tma{spectra}{} \citep[e.g.,][]{Sorai2001, Johansson1998, Heikkila1999, Israel2003, Bolatto2005, Kim2004, Kim2006, Nikolic2007, Pineda2008, MizunoY2010}. Some authors \citep[e.g.,][]{Bolatto2005, Nikolic2007} suggest two-component model especially for molecular clouds closely associated with HII regions. Most of these studies \tma{target }{targeted} \tma{outstanding }{extraordinary} HII regions \tma{and }{so} the \tma{number of}{} \tma{samples }{sample} is limited. \citet{Minamidani2008} have carried out $^{12}$CO($J=3-2$) observations of 6 GMCs, including one Type I, two Type II, and three Type III GMCs, in the LMC with the ASTE 10m telescope at a spatial resolution of 5 pc, and identified 32 molecular clumps. These data were combined with available $^{12}$CO($J=1-0$) and $^{13}$CO($J=3-2$) data and compared with LVG calculations for 13 clumps. The results show that these clumps range from cool ($\sim$ 10 -- 30K) to warm ($>$ 30 -- 200K) in kinetic temperature, and warm clumps range from less dense ($\sim$ 10$^3$ cm$^{-3}$) to dense ($\sim$ 10$^{3.5}$ -- 10$^5$ cm$^{-3}$) in density, whereas only lower limits in kinetic temperature were obtained in the warm clumps. Most recently, \citet{MizunoY2010} have made $^{12}$CO($J=4-3$) observations of the N159 region with the NANTEN2 4m sub-millimeter telescope. These data were used in LVG analysis combined with $^{12}$CO($J=1-0$), ($J=2-1$), ($J=3-2$), and ($J=7-6$) as well as the isotope transitions of $^{13}$CO($J=1-0$), ($J=2-1$), ($J=3-2$), and ($J=4-3$). The kinetic temperatures and densities were found to be $\sim$ 70 -- 80K and $\sim$ 3$\times 10^3$ cm$^{-3}$ in N159W and N159E, and $\sim$30K and $\sim$ 1.6$\times 10^3$ cm$^{-3}$ in N159S, indicating that an analysis including higher-$J$ transitions of both $^{12}$CO and $^{13}$CO can better constrain kinetic temperature and density. 

In the present study, we aim \tma{at determining }{to determine} temperature and density of \tma{molecular gas }{molecular hydrogen ,H$_2$, gas} in all three GMC types, Type I, II, and III, with higher accuracy by combining the $^{13}$CO($J=3-2$) data newly obtained by using the ASTE telescope and $^{12}$CO($J=3-2$) and $^{13}$CO($J=1-0$) data obtained with the ASTE and the SEST telescopes, respectively. \tma{These }{All our} data sets \tma{are all }{were} convolved to the same resolution of 45$\arcsec$, corresponding to $\sim$ 10pc at a distance of the LMC, 50kpc \citep{Feast1999}, and the large velocity gradient (LVG) calculations are performed to estimate the line intensities.

\tma{This paper is organized as follows:}{}Section 2 describes observations of $^{13}$CO($J=3-2$) transition. Section 3 shows the observational results and describes ancillary data sets. Section 4 shows data analysis, and in Section 5, we discuss the physical properties of clumps and evolutionary sequence of GMCs. In Section 6, we present the summary.

\section{$^{13}$CO($J=3-2$) observations}

\subsection{Selection of molecular clumps}

The present targets are chosen from the 32 molecular clumps identified by the $^{12}$CO($J=3-2$) observations of the six GMCs in the LMC with the ASTE 10m telescope at a spatial resolution of 5pc \citep{Minamidani2008}. The six GMCs were chosen from the NANTEN catalog of $^{12}$CO($J=1-0$) GMCs compiled by \citet{Fukui2008}, and they were three Type III GMCs, two Type II GMCs, and one Type I GMC. 

In the present study, we selected 9 molecular clumps in the Type III GMCs, LMC N J0538-6904 (the \tma{30Dor }{30 Doradus} region; 4 clumps), and LMC N J0540-7008 (the N159 region; 3clumps), Type II GMC, LMC N J0532-7114 (the N206D region; 1 clump), and Type I GMC, LMC N J0547-7041 (the GMC225 region; 1 clump). The observed clumps and their coordinates are listed in Table 1. Hereafter, the region names, which are parentheses above or column (4) in Table 1, and the numbers of clumps are used to identify clumps, as same as that in \citet{Minamidani2008}. Among 9 molecular clumps, four clumps are in the \tma{30Dor }{30 Doradus} region, \tma{30Dor No.1, No.2, No.3, and No.4, }{30 Doradus No. 1, No. 2, No. 3, and No. 4,} three clumps are in the N159 region, \tma{N159 No.1, No.2, and No.4, }{N159 No. 1, No. 2, and No. 4,} one clump is in the N206D region, \tma{N206D No.1, }{N206D No. 1,} and one clump is in the GMC225 region, \tma{GMC225 No.1}{GMC225 No. 1}.

\subsection{$^{13}$CO($J=3-2$) observations}

The observations of $^{13}$CO($J=3-2$) transition at 330.587960GHz were made with the ASTE 10-m telescope at Pampa la Bola in Chile in September, 2006. In this term, a single cartridge type double-side-band (DSB) SIS receiver, SC345 \citep{Kohno2005}, was installed and the XF-type digital autocorrelator \citep{Sorai2000} was operated in the wideband mode, providing a bandwidth of 512 MHz with 1024 channels. These correspond to a velocity coverage of 450 km s$^{-1}$ and the resolution of 0.45 km s$^{-1}$ at 330GHz. The half power beam width was 23$\arcsec$ at 330GHz, and this corresponds to 5.6pc at 50kpc. We \tma{have}{} made 3$\times$3 points mapping observations with position switching method with a 20$\arcsec$ grid spacing toward the peaks of $^{12}$CO($J=3-2$) clumps detected by \citet{Minamidani2008}. Figure 1 shows the observed positions of the peaks of the $^{12}$CO($J=3-2$) clumps named \tma{30Dor No.1, No.2, No.3, No.4, N159 No.1, No.2, No.4, N206D No.1, and GMC225 No.1. }{30 Doradus No. 1, No. 2, No. 3, No. 4, N159 No. 1, No. 2, No. 4, N206D No. 1, and GMC225 No. 1.} The typical system temperature including the atmosphere was 260K in DSB. The pointing accuracy was measured to be better than 5$\arcsec$ in peak to peak by periodically  observing a CO point source R Dor in the $^{12}$CO($J=3-2$) transition every 2 hours. The spectral intensities were calibrated by employing the standard room-temperature chopper-wheel method. We observed Ori-KL and M17SW in the $^{13}$CO($J=3-2$) transition to check the stability of the intensity calibration, and the intensity variation during these observations was less than 12\%. 

\tma{}{The observed antenna temperature $T\mathrm{_A^*(OBS)}$ varies with the side-band ratio ($SBR$) or the image rejection ratio ($IRR$): 
\begin{displaymath}
	T\mathrm{_A^*(OBS)} = \frac{T\mathrm{_A^*(SSB)}}{1 + \frac{1}{IRR}}
\end{displaymath}
where $T\mathrm{_A^*(SSB)}$ is the antenna temperature in the single side-band (SSB). We measured the antenna temperature, $T\mathrm{_A^*(OBS)}$, and image rejection ratio ($IRR$), simulteneously, with the newly installed 2SB receiver, CATS345 \citep{Ezawa2008, Inoue2008}, to the ASTE telescope in 2008, and derived the SSB antenna temperatures, $T\mathrm{_A^*(SSB)}$, of IRC+10216 [R.A.(1950) = $\mathrm{09^h45^m14.8^s}$, Dec.(1950) = +13\degr 30\arcmin 40\arcsec ] and N159W [R.A. (B1950) = $\mathrm{05^h40^m03.7^s}$, Dec. (B1950) = -69\degr 47\arcmin 00.0\arcsec ]. From the observation of IRC+10216, the observed velocity integrated intensity, $I.I.$($T\mathrm{_A^*(OBS)}$), was 56$\pm$4 K km s$^{-1}$, and the image rejection ratio, $IRR$, was 6$\pm$2 ($\sim$8 dB). This $IRR$ was consistent with the results of the laboratory evaluation \citep{Inoue2008}. The $I.I.$($T\mathrm{_A^*(SSB)}$) was estimated to be 65$\pm$6 K km s$^{-1}$. This value was compared with the data taken by CSO \citep{Wang1994}, and main beam efficiency at that moment was derived to be 0.70$\pm$0.06. From the observation of N159W, the observed antenna temperature, $T\mathrm{_A^*(OBS)}$, was 1.83$\pm$0.04 K, and the image rejection ratio, $IRR$, was 4.7$\pm$0.4 ($\sim$7 dB). This $IRR$ was consistent with the results of the laboratory evaluation (Inoue et al. 2008).  The SSB antenna temperature, $T\mathrm{_A^*(SSB)}$, was estimated to be 2.22$\pm$0.06 K. We derive the main beam temperature of N159W to be 3.2$\pm$0.3 K. }%
We scaled the observational data in 2006 to be consistent with this value. The achieved noise level ranges from 0.04 K to 0.18 K in the main beam temperature scale. These observations were made remotely from NAOJ and NRO in Japan, by using the network observation system N-COSMOS3 developed by NAOJ \citep{Kamazaki2005}.

\section{Results \& Data sets}

\subsection{$^{13}$CO($J=3-2$)}

In Figure 2, $^{13}$CO($J=3-2$) profile maps of each observed peak are presented with $^{12}$CO($J=3-2$) spectra \citep{Minamidani2008}. The $^{13}$CO($J=3-2$) intensities are 3 -- 12 times lower than those of $^{12}$CO($J=3-2$). The peak velocities are consistent with those of $^{12}$CO($J=3-2$), and the line widths are factor of 0.7 -- 1.0 smaller than those of $^{12}$CO($J=3-2$). Their line parameters at the peak positions of $^{12}$CO($J=3-2$) are summarized in Table 1. Detailed descriptions explaining the characteristics of each observed $^{12}$CO($J=3-2$) peak are given in Appendix.

\subsection{$^{12}$CO($J=3-2$)}

We use the $^{12}$CO($J=3-2$) data published by \citet{Minamidani2008}, which were taken by the ASTE telescope with an angular resolution of 22$\arcsec$ . 
\tma{}{We measured the antenna temperature, $T\mathrm{_A^*(OBS)}$, and image rejection ratio ($IRR$), simulteneously, with the newly installed 2SB receiver, CATS345 \citep{Ezawa2008, Inoue2008}, to the ASTE telescope in 2008, and derive the SSB antenna temperatures, $T\mathrm{_A^*(SSB)}$, of IRC+10216 [R.A.(1950) = $\mathrm{09^h45^m14.8^s}$, Dec.(1950) = +13\degr 30\arcmin 40\arcsec ] and N159W [R.A. (B1950) = $\mathrm{05^h40^m03.7^s}$, Dec. (B1950) = -69\degr 47\arcmin 00.0\arcsec ]. From the observation of IRC+10216, the observed antenna temperature, $T\mathrm{_A^*(OBS)}$, was 18.8$\pm$0.7 K, and the image rejection ratio, $IRR$, was 22$\pm$3 ($\sim$13 dB). This $IRR$ was consistent with the results of the laboratory evaluation \citep{Inoue2008}. The SSB antenna temperature, $T\mathrm{_A^*(SSB)}$, was estimated to be 19.6$\pm$0.8 K. This value was compared with the data taken by CSO \citep{Wang1994}, and main beam efficiency at that moment was derived to be 0.60$\pm$0.02. From the observation of N159W, observed antenna temperature, $T\mathrm{_A^*(OBS)}$, was 7.6$\pm$0.2 K, and image rejection ratio, $IRR$, was 9$\pm$2 ($\sim$10 dB). This $IRR$ was consistent with the results of the laboratory evaluation (Inoue et al. 2008). The SSB antenna temperature, $T\mathrm{_A^*(SSB)}$, was estimated to be 8.4$\pm$0.2 K. We derive the main beam temperature of N159W to be 13.9$\pm$0.7 K.} We scaled the observational data taken in 2004 \citep{Minamidani2008} to be consistent with this value.

\subsection{$^{13}$CO($J=1-0$)}

We use the $^{13}$CO($J=1-0$) intensity of the 30 Doradus and N159 regions published by \citet{Johansson1998}, and the N206D and GMC225 regions published by \citet{Minamidani2008}. These data are all taken by the SEST telescope, and the angular resolution was 45$\arcsec$.

\section{Data analysis}

\subsection{Derivation of line intensity ratios}

The spatial resolutions of the present CO data vary depending on the telescope and frequency. The angular resolutions of $^{12}$CO($J=3-2$) and $^{13}$CO($J=3-2$) data observed by the ASTE telescope are 22$\arcsec$ and 23$\arcsec$, respectively, \tma{although }{while} that of $^{13}$CO($J=1-0$) data observed by the SEST telescope are 45$\arcsec$. These correspond to $\sim$ 5 pc and 10 pc, respectively, at 50kpc. We have convolved the $^{12}$CO($J=3-2$) and $^{13}$CO($J=3-2$) data into the 45$\arcsec$ beam with a Gaussian \tma{kernel }{smoothing function} in order to derive physical properties of clumps whose sizes are around 10 pc.

We \tma{made Gaussian curve fits }{fitted Gaussians} to each of the $^{12}$CO($J=3-2$), $^{13}$CO($J=3-2$), and $^{13}$CO($J=1-0$) spectra  with a single spectral peak. We derived the temperature peaks of the Gaussian curves, and the results are summarized in Table 2. The ratios of $^{12}$CO($J=3-2$) to $^{13}$CO($J=3-2$) (hereafter, $R^{12/13}_{3-2}$) and $^{13}$CO($J=3-2$) to $^{13}$CO($J=1-0$) (hereafter, $R^{13}_{3-2/1-0}$) are derived as the ratios of the peak values \tma{in the }{on a} main beam temperature scale. The errors of $R^{12/13}_{3-2}$ and $R^{13}_{3-2/1-0}$ are estimated to be 20 -- 27\% and 25 -- 30\%, respectively. A summary of these main beam temperature ratios and their accuracies are also presented in Table 2. These ratios will be compared with numerical calculations of radiative transfer in the LVG approximation to derive constraints on density and temperature \tma{as follows}{}.

\subsection{LVG analysis}

\subsubsection{Calculations of a LVG model}

In order to estimate physical properties of molecular gas, we have performed an LVG analysis \citep{Goldreich1974} of the CO rotational transitions in the same way as described in \citet{Minamidani2008}. The LVG radiative transfer code simulates a spherically symmetric cloud of uniform density and temperature with a spherically symmetric velocity gradient proportional to the radius, and employs a Castor's escape probability formalism \citep{Castor1970}. It solves the equations of statistical equilibrium for the fractional population of CO rotational levels at each density and temperature\tma{.}{, because the level populations of CO are determined by both density and kinetic temperature of molecular hydrogen.} It includes the lowest 40 rotational levels of the ground vibrational level and use the \tma{Einstein's }{Einstein} A coefficient and H$_2$ collisional impact rate coefficients obtained from the \tma{Liden }{Leiden} Atomic and Molecular Database \citep[LAMDA;][]{Schoier2005}. 

We performed calculations of the fractional populations of the lowest 40 rotational levels of $^{12}$CO and $^{13}$CO in the ground vibrational state over a kinetic temperature range of $T_{\mathrm{kin}}$ = 5 -- 200K and a density range of $n$(H$_2$) = 10 -- 10$^6$ cm$^{-3}$. We assumed that the fractional abundance of CO to H$_2$ and the abundance ratio of $^{12}$CO to $^{13}$CO are 1.6$\times 10^{-5}$ and 50, respectively \citep{MizunoY2010, Blake1987}. 

Figure 3 shows the general behavior of the loci of constant $R^{13}_{3-2/1-0}$ and constant $R^{12/13}_{3-2}$ in the density-temperature plane. The $R^{13}_{3-2/1-0}$ is the ratio of transitions which have different critical densities, and traces excitation of molecular gas. The $R^{12/13}_{3-2}$ is the ratio of transitions which have same critical densities and trace different column densities, and thus this ratio is a good tracer of column density and consequently volume density in wide temperature range. \tma{It is recognized that the }{The} combination of the two line ratios is nearly "orthogonal" \tma{in the plane}{}, \tma{and }{so} the density and kinetic temperature are well constrained. This is different from the case of \citet{Minamidani2008}. \citet{Minamidani2008} used the combination of the ratios of $^{12}$CO($J=3-2$) to $^{12}$CO($J=1-0$) (here after, $R^{12}_{3-2/1-0}$) and $^{12}$CO($J=1-0$) to $^{13}$CO($J=1-0$) (here after, $R^{12/13}_{1-0}$), which does not constrain well the physical parameters for densities higher than 10$^4$ cm$^{-3}$ \citep{Minamidani2008}.

\subsubsection{Results of the LVG analysis}

We summarize the input parameters for the 8 clumps in columns (3) -- (5) of Table 3. The higher transition data have been convolved into the 45$\arcsec$ beam with a Gaussian kernel as described in section 4.1. The velocity gradient, $dv/dr$, of each clump is derived from $^{12}$CO($J=3-2$) data \citep{Minamidani2008}.

Figure 4 shows the results of the LVG analysis for the 8 clumps. The horizontal axis is molecular hydrogen density, $n$(H$_2$), and the vertical axis is the gas kinetic temperature, $T_{\mathrm{kin}}$. Solid lines represent $R^{13}_{3-2/1-0}$ and dashed lines represent $R^{12/13}_{3-2}$. \tma{Hatched }{The hatched} areas indicate the \tma{overlapped }{overlap} regions of these two ratios within the errors. In Table 3, we present the estimated densities and kinetic temperatures of the clumps. 

The derived density and kinetic temperature are well determined compared to the previous work based on the combination of $R^{12}_{3-2/1-0}$ and $R^{12/13}_{1-0}$ \citep{Minamidani2008}, especially for the clumps in the warm Type III GMCs. The results for the three clumps in the N159 region show a good agreement with the results of \citet{MizunoY2010} who employed the calculations using high-$J$ transitions of $J=4-3$.

Figure 5 summarizes the derived kinetic temperatures \tma{}{($T\mathrm{_{kin}}$)} and densities \tma{}{($n$(H$_2$))} of all 8 clumps. \tma{They }{These} range from 15 to 200 K in kinetic temperature and from 8$\times 10^2$ to 7$\times 10^3$ cm$^{-3}$ in density. We found clumps in the \tma{30Dor}{30 Doradus} and N159 regions are warm and dense, except for N159 \tma{No.4}{No. 4} (N159S), which is less dense and has \tma{}{an} intermediate kinetic temperature. The clump in N206D, N206D \tma{No.1}{No. 1}, is dense and has \tma{also }{an} intermediate kinetic temperature. The clump in GMC225, GMC225 \tma{No.1}{No. 1}, is less dense and cold. The derived kinetic temperature and density denote the same tendency as indicated by Minamidani et al. (2008) with improved accuracies. For example, the kinetic temperature and density of the N159 \tma{No.1 }{No. 1} (N159W) clump are determined with the accuracies of 39\% and 31\%, respectively, in the present work, although, in the previous study \citep{Minamidani2008}, only lower limit of the kinetic temperature (30K) and the two orders of magnitude uncertainty of density (3$\times 10^3$ -- 8$\times 10^5$ cm$^{-3}$) were suggested for this clump.

\subsubsection{Discussion on the LVG results}

There are two distinctive aspects in the present work. One is the continuous coverage of molecular clump samples in the all types of GMCs, Type I, II, and III, which are interpreted as an evolutionary sequence of GMCs (Kawamura et al. 2009). The other is the combinations of line ratios used in the LVG analysis. The $^{13}$CO($J=3-2$), $^{12}$CO($J=3-2$), and $^{13}$CO($J=1-0$) transitions are used for the analysis and the $^{12}$CO($J=1-0$) transition is not included in this work. This improves the accuracies of kinetic temperature and density of molecular clumps as described by Mizuno et al. (2010).

We also performed LVG calculations with the parameters used by \citet{Minamidani2008}, which are $X$(CO) = 3.0$\times 10^{-6}$ and abundance ratio of $^{12}$CO to $^{13}$CO of 20, 25, and 30. The kinetic temperatures and densities of clumps estimated from $R^{12/13}_{3-2}$ and $R^{13}_{3-2/1-0}$ with these parameters were well determined. This indicates that the combinations of line intensity ratios used in this paper, $R^{12/13}_{3-2}$ and $R^{13}_{3-2/1-0}$, are suitable to determine physical properties of molecular clumps in wide ranges.

The N159 region\tma{,}{ is} the most active site of high mass star formation in the Large Magellanic Cloud\tma{,}{. This} has been observed in various molecular transitions by various telescopes located in the southern hemisphere, and many authors paid their efforts to determine the physical properties and chemical compositions via LVG analyses \citep[e.g.,][]{MizunoY2010, Pineda2008, Minamidani2008}. In Table 4, some of the recent LVG results of the N159 region are summarized. The kinetic temperature and density derived in this work are consistent within errors with that derived by \citet{MizunoY2010} and \citet{Pineda2008}, which are determined with high accuracies based on data of higher transitions of $J=4-3$ and $J=7-6$.

\subsection{Comparison of line intensity ratios and physical properties with H$\alpha$ flux}

We use the H$\alpha$ data \citep{Kim1999} toward the present clouds using the method given in Appendix B in \citet{Minamidani2008}. The typical background level of H$\alpha$ flux is $\sim$ 10$^{-12}$ ergs s$^{-1}$ cm$^{-2}$ at the 40$\arcsec$ scale, which is a pixel scale of the H$\alpha$ data. These data were regridded into the $^{13}$CO($J=3-2$) data grids. The fluxes toward each $^{13}$CO($J=3-2$) peak are listed in Tables 2 and 3.

Figure 6 shows plots of $R^{12/13}_{3-2}$ and $R^{13}_{3-2/1-0}$ as functions of H$\alpha$ flux. It is clear that $R^{13}_{3-2/1-0}$ is well correlated with the H$\alpha$ flux with a correlation coefficient of 0.98, and there is no clear correlation between $R^{12/13}_{3-2}$ and H$\alpha$ flux.

Figure 7 shows plots of \tma{density }{molecular hydrogen densities, $n$(H$_2$),} and kinetic \tma{temperature }{temperatures, $T\mathrm{_{kin}}$,} as functions of H$\alpha$ flux. \tma{Dense }{For dense} clumps \tma{}{these} are distributed \tma{in }{over} a wide range of H$\alpha$ flux (10$^{-12}$ -- 10$^{-10}$ ergs s$^{-1}$ cm$^{-2}$), although lower density clumps are located only where H$\alpha$ flux is weak ($\sim$ 10$^{-12}$ ergs s$^{-1}$ cm$^{-2}$). The \tma{kinetic temperature }{$T\mathrm{_{kin}}$} of clumps \tma{is }{are} well correlated with the H$\alpha$ flux with a correlation coefficient of 0.81. We note that the size scale here is 10pc and the good correlation may not hold at smaller scales where local extinction of H$\alpha$ becomes important \citep[e.g.,][]{MizunoY2010}. These results suggest that far-ultraviolet (FUV) photons heat molecular gas whose density is 10$^3$ -- 10$^4$ cm$^{-3}$. This will be discussed in the following section.

\section{Discussions}

\subsection{Evolution\tma{}{s} of GMC and clump}

The results of our LVG analysis show that the clump kinetic \tma{temperature }{temperatures, $T\mathrm{_{kin}}$,} \tma{distributes }{range} from cool ($\sim$ 15K) to warm ($\sim$ 200K) and the \tma{density }{densities, $n$(H$_2$),} from less dense ($\sim$ 8$\times 10^2$ cm$^{-3}$) to dense ($\sim$ 7$\times 10^3$ cm$^{-3}$). These large variations in the physical properties reflect the different characteristics of the GMCs according to their \tma{types }{Types} I, II, and III\tma{}{, where Type I shows no signs of massive star formation, Type II associated with only small HII regions, and Type III associated with both HII regions and young stellar clusters, and these Types are interpreted as an evolutionary sequence \citep{Kawamura2009}}. A clump in \tma{}{a} Type I GMC, GMC225 No. 1, has the lowest density (\tma{}{$n$(H$_2$)} $\sim$ 0.99$\times 10^2$ cm$^{-3}$ ) and lowest temperature (\tma{}{$T\mathrm{_{kin}}$} $\sim$ 25K). A clump in Type II GMC, N206D No. 1, is a dense (\tma{}{$n$(H$_2$)} $\sim$ 3.2$\times 10^3$ cm$^{-3}$) clump and has intermediate temperature (\tma{}{$T\mathrm{_{kin}}$} $\sim$ 42K). Clumps in Type III GMCs, except for N159 No. 4 (N159S) clump, are dense (\tma{}{$n$(H$_2$)} $\sim$ 4$\times 10^3$ cm$^{-3}$) and warm (\tma{}{$T\mathrm{_{kin}}$} $>$ 50K) clumps. The physical properties are generally correlated with the star formation activity of GMCs. The \tma{kinetic temperature }{$T\mathrm{_{kin}}$} and \tma{density }{$n$(H$_2$)} increase with the evolution of GMCs from Type I to III. This trend becomes clearer with the present determination of \tma{kinetic temperature }{$T\mathrm{_{kin}}$} and \tma{density }{$n$(H$_2$)} than that of the previous study \citep{Minamidani2008}, although the number of clump samples is smaller than that.

The N159 No. 4 (N159S) clump is a part of \tma{}{a} Type III GMC, LMC N J0540-7008 \citep{Kawamura2009}. This GMC is quite large and elongated from north to south \citep[275pc $\times$ 53pc, P.A. = 87\degr;][]{Fukui2008}. Young clusters and large HII regions are associated with the northern part of this GMC, and in the southern part, where N159 No. 4 (N159S) is located, young clusters and large HII regions are not associated with. The N159 No. 4 (N159S) clump is less dense (\tma{}{$n$(H$_2$)} $\sim$ 1.5$\times 10^3$ cm$^{-3}$) and has intermediate temperature (\tma{}{$T\mathrm{_{kin}}$} $\sim$ 39K). These properties are quite similar to that of the clumps in Type I or II GMCs, as discussed by \citet{Minamidani2008} and \citet{MizunoY2010}. Comparison of the properties of molecular clumps with signs of star formation activities, such as young star clusters and H$\alpha$ emission, at high spatial resolution will establish the classification of molecular clumps, whose sizes are $\sim$ 7 pc \citep{Minamidani2008}. Because the size of clusters in the LMC are distributed in a range from 0.1 -- 10 pc \citep{Hunter2003}, which is smaller/similar size than the typical size of molecular clumps ($\sim$ 7 pc), it is important to compare molecular clumps with signs of star/clustar formation to understand the evolution of molecular clumps leading to the star/cluster formation.

\subsection{Heating of the molecular gas in the LMC}

As shown in the previous sections, the $R^{13}_{3-2/1-0}$ is well correlated with H$\alpha$ flux at a 10 pc scale, and the molecular gas kinetic temperatures \tma{}{($T\mathrm{_{kin}}$)} are also well correlated with H$_{\alpha}$ flux. These present findings suggest that the heating of molecular gas whose densities are 10$^3$ -- 10$^4$ cm$^{-3}$ may be dominated by far-ultraviolet (FUV) photons. The intense FUV field controls the physical and chemical processes in the ISM such as formation and destruction of molecules as well as ionization. These regions have been modeled as photo-dissociation regions (PDRs) or photon-dominated regions (PDRs) \citep[e.g.,][and references their in]{Tielens1985, Kaufman1999, Rollig2007}. 

The FUV flux ($G_0$) is estimated as 3500 in the 30 Doradus region \citep{Bolatto1999, Poglitsch1995, Werner1978, Israel1979} and 300 for the N159 region \citep{Bolatto1999, Israel1996, Israel1979}\tma{,}{.} \tma{and the }{The} gas density is estimated \tma{as }{to be} (1 -- 5)$\times 10^3$ cm$^{-3}$ for these regions\tma{ in the present is work}{}. The PDR surface temperature \tma{can be read as }{is} $\sim$ 400 K for the \tma{30Doradus }{30 Doradus} region and $\sim$ 200 K for the N159 region from Figure 1 of \citet{Kaufman1999}. The PDR gas temperature is relatively constant from the cloud surface to a depth where either heating or cooling changes significantly. The heating is generally dominated by the grain photoelectric heating \citep{Kaufman1999}, and then dust attenuation of FUV flux controls the thermal structure, where the typical size scale is 0.1 -- 2 pc. These suggest that the effect of FUV heating of molecular gas seems to be local and direct phenomena. 

The warm region can become larger under low-metallicity or high gas-to-dust ratio environments. These temperatures are basically consistent with the kinetic temperatures \tma{}{($T\mathrm{_{kin}}$)} of the warm clumps in the present sample if beam dilution by the present resolution, 10 pc, is taken into account.

\subsection{Substructures inferred by the $^{13}$CO($J=3-2$) observations}

The $^{13}$CO($J=3-2$) intensities are 3 -- 12 times lower than those of $^{12}$CO($J=3-2$) transition, and the line intensity ratios of $^{12}$CO($J=3-2$) to $^{13}$CO($J=3-2$), $R^{12/13}_{3-2}$\tma{,}{.} \tma{}{These} vary not only from clump to clump but also inside of each clump. In the some clumps, clear two velocity components are detected in the $^{13}$CO($J=3-2$) transition. These suggest some internal structures.

\citet{MizunoY2010} showed that the molecular distribution of the N159 No. 1 (N159W) clump in the $^{12}$CO($J=3-2$) transition is similar to that of the $\eta$ Carinae northern cloud in the $^{12}$CO($J=1-0$) emission line smoothed to 5 pc resolution. The original data of $\eta$ Carinae northern cloud has a 2 pc resolution and several substructures are identified \citep{Yonekura2005}. This supports the existence of internal structures inside of $\sim$ 5 pc scale molecular clumps, and indicates that these internal structures can be resolved in the CO transitions with high spatial resolution observations.

\tma{The subsequent }{Subsequent} observations of $\eta$ Carinae northern cloud using a high density tracer such as H$^{13}$CO$^{+}$($J=1-0$) at a high spatial resolution, \tma{}{resulted in the detection of} high density molecular cores whose sizes are less than 1 pc \tma{were detected}{}\citep{Yonekura2005}. In the LMC, some \tma{pioneering }{initial} interferometric observations of high density tracers, such as HCO$^{+}$ and NH$_3$, were made \tma{by }{with} ATCA, \tma{whose }{using} resolutions \tma{were }{of} $\sim$ 6$\arcsec$ -- 19$\arcsec$ corresponding to $\sim$ 1.4 -- 5 pc at 50 kpc \citep[e.g.,][]{Wong2006, Ott2008, Ott2010}. \citet{Ott2008} detected two peaks in the HCO$^{+}$($J=1-0$) transition in the N159W region. Further systematic detailed observations using high density tracers at higher resolution, with ALMA for instance, should be important for probing the initial conditions of star/cluster formation.

\section{Summary}

\tma{We summarize the results as follows.}{}

\tma{1)}{} We have made 3 $\times$ 3 points mapping observations in the $^{13}$CO($J=3-2$) transition at 330 GHz using the ASTE 10m telescope toward 9 peaks of $^{12}$CO($J=3-2$) clumps, which cover all types of GMCs, Type I, II, and III, which are interpreted as an evolutionary sequences of GMCs. We have detected $^{13}$CO($J=3-2$) emission from all peaks and their intensities are 3 -- 12 times lower than those of $^{12}$CO($J=3-2$). \tma{}{From this, : }%
\tma{2) }{}We have derived the intensity ratios of $^{12}$CO($J=3-2$) to $^{13}$CO($J=3-2$), $R^{12/13}_{3-2}$, and $^{13}$CO($J=3-2$) to $^{13}$CO($J=1-0$), $R^{13}_{3-2/1-0}$, at 45$\arcsec$ resolution, and have compared these results with the LVG radiative transfer calculations in order to estimate kinetic \tma{temperature }{temperatures, $T\mathrm{_{kin}}$}, and \tma{density }{densities, $n$(H$_2$),} of the 8 clumps. The clumps show \tma{temperature of }{$T\mathrm{_{kin}}$ =} 15 -- 200 K and \tma{density of }{$n$(H$_2$) =} 8$\times 10^2$ -- 7$\times 10^3$ cm$^{-3}$.%
\tma{3)}{} The H$\alpha$ flux toward these clumps is well correlated with the $^{13}$CO($J=3-2$)/$^{13}$CO($J=1-0$) ratio, $R^{13}_{3-2/1-0}$, and with the kinetic temperatures, \tma{}{$T\mathrm{_{kin}}$,} of the clumps. Dense clumps are distributed in a wide range of H$\alpha$ flux (10$^{-12}$ -- 10$^{-10}$ ergs s$^{-1}$ cm$^{-2}$), although lower density clumps are located only where H$\alpha$ flux is weak ($\sim$ 10$^{-12}$ ergs s$^{-1}$ cm$^{-2}$).%
\tma{4)}{} We found \tmb{}{that} clumps in the \tma{30Dor }{30 Doradus} and N159 regions (Type III GMCs) are warm and dense, except for N159 No. 4 (N159S), which is less dense and has intermediate kinetic temperature. The clump in N206D (Type II GMC), N206D No. 1, is dense and has also intermediate kinetic temperature. The clump in GMC225 (Type I GMC), GMC225 No. 1, is less dense and cold. We suggest that differences of these clump properties largely reflect an evolutionary sequence of GMCs and molecular clumps. The kinetic temperatures \tma{}{($T\mathrm{_{kin}}$)} and densities \tma{}{($n$(H$_2$))} of molecular clumps increase generally with the evolution of GMCs and molecular clumps.%
\tma{5)}{} The $R^{13}_{3-2/1-0}$ and kinetic temperature\tma{}{s ($T\mathrm{_{kin}}$)} are well correlated with H$\alpha$ flux, suggesting that the heating of molecular gas whose densities \tma{}{($n$(H$_2$)) =} 10$^3$ -- 10$^4$ cm$^{-3}$ is dominated by FUV photons. The calculations of PDR models are consistent with this suggestion.



\acknowledgments

A part of this study was financially supported by MEXT Grant-in-Aid for Specially Promoted Research (20001003). T.M. was supported by JSPS Research Fellowships for Young Scientists. M.R. wishes to acknowledge support from FONDECYT(CHILE) grant No1080335 and the Chilean {\sl Center for Astrophysics} FONDAP No. 15010003.

The ASTE project is \tmb{driven }{managed} by Nobeyama Radio Observatory (NRO), a branch of National Astronomical Observatory of Japan (NAOJ), in collaboration with Uversity of Chile, and Japanese institutes including University of Tokyo, Nagoya University, Osaka Prefecture University, Ibaraki University, and Hokkaido University. Observations with ASTE were carried out remotely from Japan by using NTTfs GEMnet2 and its partnet R\&E (Research \& Education) networks, which are based on AccessNova collaboration of University of Chile, NTT Laboratories, and NAOJ.



{\it Facilities:} \facility{ASTE (SC345-MAC)}, \facility{SEST}.



\appendix

\section{\tmb{Detaild }{Detailed} descriptions of each clump}

Detailed descriptions of characteristics of each clump are presented in the following.

\paragraph{30 Doradus No. 1 (30Dor-10) (Figure 2a)}
The \tmb{achieved }{measured} noise level of $^{13}$CO($J=3-2$) spectra in this region is 0.10 K rms at 0.45 km s$^{-1}$ velocity resolution. The spatial extent and the velocity range of the $^{13}$CO($J=3-2$) emission are well correlated to those of the $^{12}$CO($J=3-2$) emission. The ratios of integrated intensities of $^{12}$CO($J=3-2$) to $^{13}$CO($J=3-2$) (hereafter $R^{12/13}_{3-2, II}$) are relatively small ($\sim$ 6) at the western position and relatively large ($\sim$ 19) at the northern position.

\paragraph{30 Doradus No. 2 (30Dor-12) (Fig. 2b)}
The \tmb{achieved }{measured} noise level of $^{13}$CO($J=3-2$) spectra in this region is 0.09 K rms at 0.45 km s$^{-1}$ velocity resolution. The spatial extent and the velocity range of $^{13}$CO($J=3-2$) emission are well correlated to those of the $^{12}$CO($J=3-2$) emission. $R^{12/13}_{3-2, II}$ at the northern position is $\sim$ 3, the smallest value among this clump, and is larger than 15 at the southern position.

\paragraph{30 Doradus No. 3 \& No. 4 (30Dor-06) (Fig. 2c)}
The southern part corresponds to No.3 clump and the northern part corresponds to No.4 clump. The \tmb{achieved }{measured} noise level of $^{13}$CO($J=3-2$) spectra in this region is 0.08 K rms at 0.45 km s$^{-1}$ velocity resolution. The spatial extent and the velocity range of the $^{13}$CO($J=3-2$) emission are well correlated to those of the $^{12}$CO($J=3-2$) emission.

\paragraph{N159 No. 1 (N159W) (Fig. 2d)}
The \tmb{achieved }{measured} noise level in this region is 0.18 K r.m.s. at 0.45 km s$^{-1}$ velocity resolution. The spatial extent and the velocity range of the $^{13}$CO($J=3-2$) emission are well correlated to those of the $^{12}$CO($J=3-2$) emission. $R^{12/13}_{3-2, II}$ shows the smallest value of $\sim$ 5 in this clump at the centeral position. At the south-west position, the $R^{12/13}_{3-2, II}$ is also small ($\sim$ 6). At the north-west position, the ratio is large of $\sim$ 15.

\paragraph{N159 No. 2 (N159E) (Fig. 2e)}
The \tmb{achieved }{measured} noise level in this region is 0.11 K r.m.s. at 0.45 km s$^{-1}$ velocity resolution. The spatial extent and the velocity range of the $^{13}$CO($J=3-2$) emission are well correlated to those of the $^{12}$CO($J=3-2$) emission. $R^{12/13}_{3-2, II}$ shows the smallest value of $\sim$ 5 in this clump at the western position and shows large value of $\sim$ 15 at the north-east position.

\paragraph{N159 No. 4 (N159S) (Fig. 2f)}
The \tmb{achieved }{measured} noise level in this region is 0.07 K r.m.s. at 0.45 km s$^{-1}$ velocity resolution. The spatial extent and the velocity range of the $^{13}$CO($J=3-2$) emission are well correlated to those of the $^{12}$CO($J=3-2$) emission. Intensities of the $^{13}$CO($J=3-2$) emission are fairly weak compared to those of the $^{12}$CO($J=3-2$) emission. $R^{12/13}_{3-2, II}$ shows the smallest value of $\sim$ 12 in this clump at the center position. Two velocity components are detected in the $^{13}$CO($J=3-2$) transition, although only one component is detected in the $^{12}$CO($J=3-2$) transition.

\paragraph{N206D No. 1 (Fig. 2g)}
The \tmb{achieved }{measured} noise level in this region is 0.09 K r.m.s. at 0.45 km s$^{-1}$ velocity resolution. The spatial extent and the velocity range of $^{13}$CO($J=3-2$) emission are well correlated to those of $^{12}$CO($J=3-2$) emission.

\paragraph{GMC225 No. 1 (Fig. 2h)}
The \tmb{achieved }{measured} noise level in this region is 0.04 K r.m.s. at 0.45 km s$^{-1}$ velocity resolution. The spatial extent and the velocity range of $^{13}$CO($J=3-2$) emission are well correlated to those of the $^{12}$CO($J=3-2$) emission. Intensities of the $^{13}$CO($J=3-2$) emission are $\sim$ 12 times weaker than those of the $^{12}$CO($J=3-2$) emission.





\clearpage



\begin{figure}[htbp]
\figurenum{1a}
\begin{center}
\includegraphics{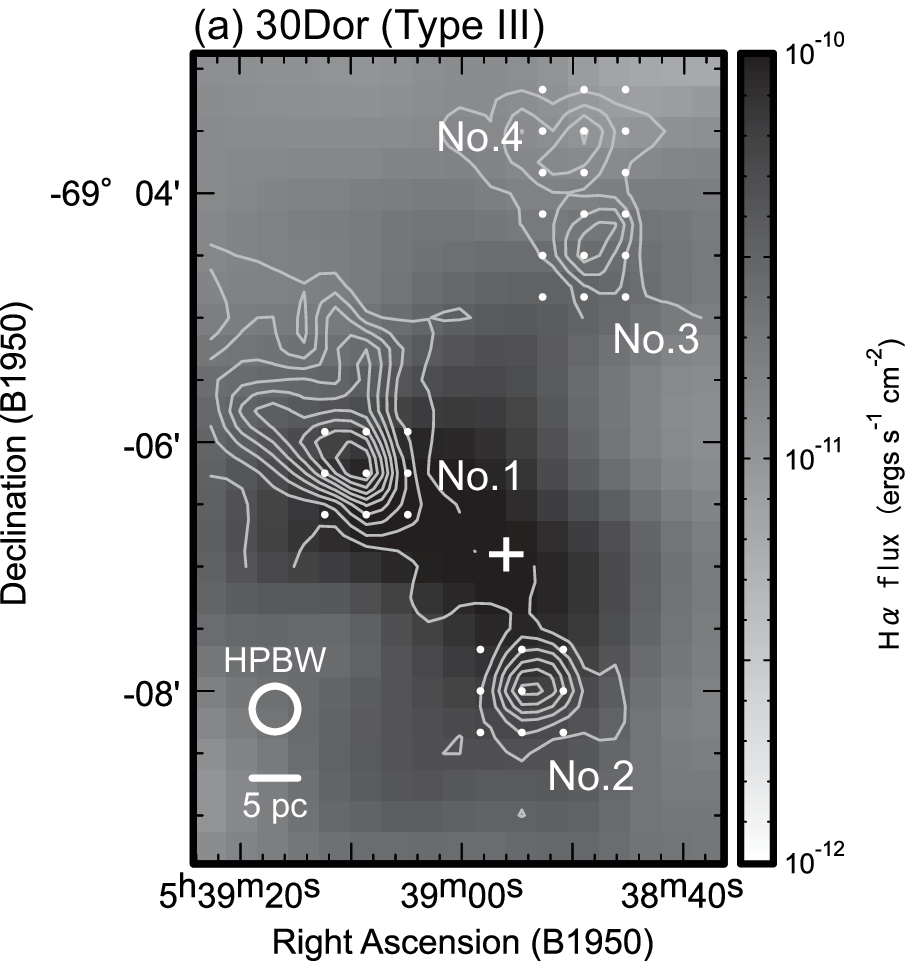}
\end{center}
\caption{Positions observed with $^{13}$CO($J=3-2$) in the (a)30 Dor, (b)N159, (c)N206D, and (d)GMC225 regions. Observed positions are indicated with dots. Contours indicate integrated intensity of $^{12}$CO($J=3-2$) and the grey scale indicates the H$\alpha$ flux (Kim et al. 1999). White crosses indicate positions of young cluster (\textless 10 Myr; SWB0). \label{fig01a}}
\end{figure}
\newpage

\begin{figure}[htbp]
\figurenum{1b}
\begin{center}
\includegraphics{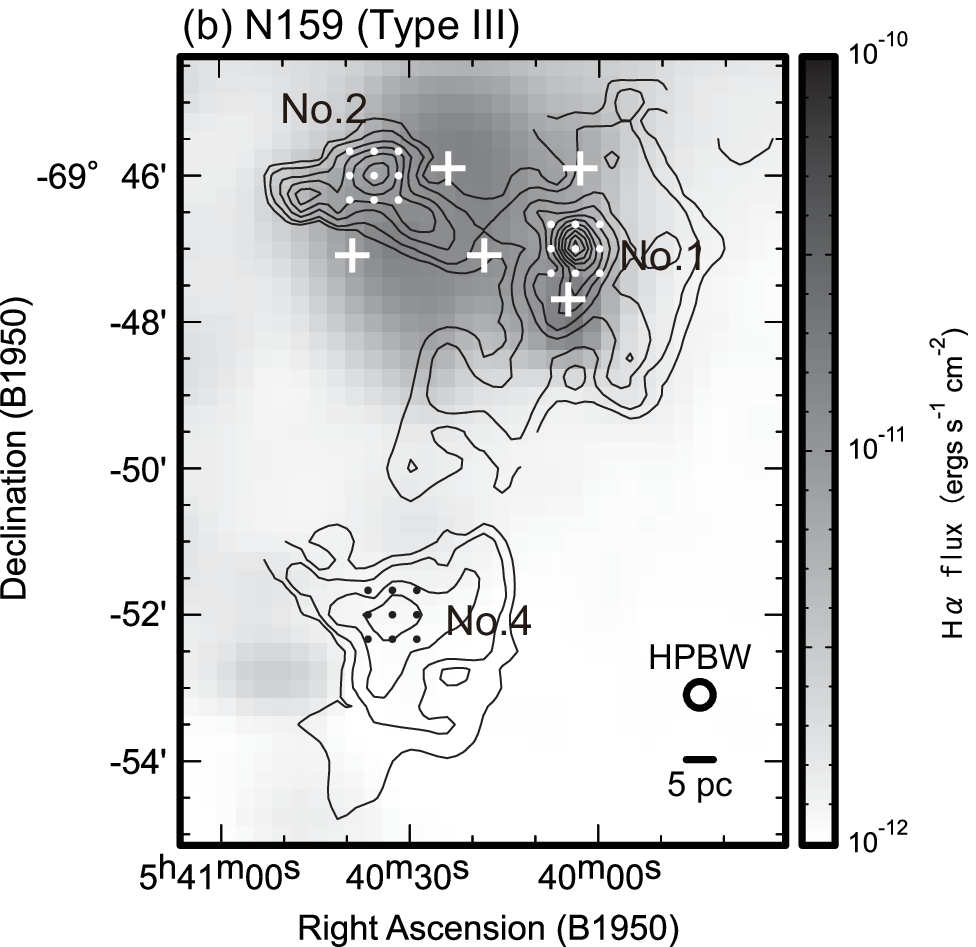}
\end{center}
\caption{ \label{fig01b}}
\end{figure}
\newpage

\begin{figure}[htbp]
\figurenum{1c}
\begin{center}
\includegraphics{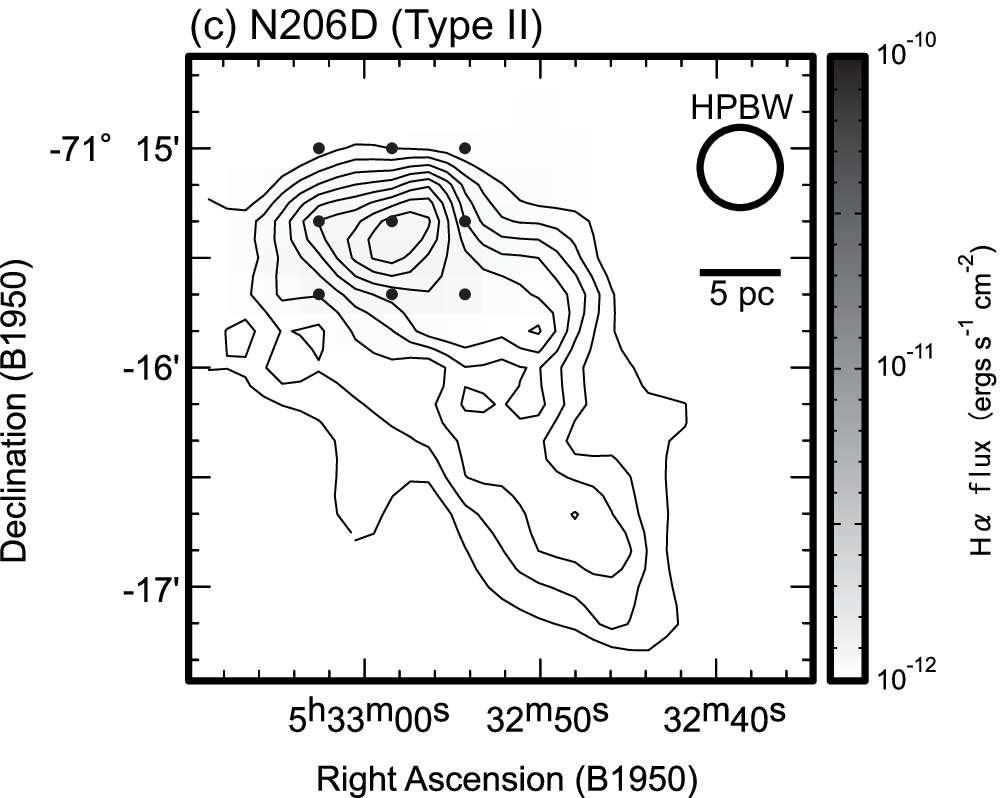}
\end{center}
\caption{ \label{fig01c}}
\end{figure}
\newpage

\begin{figure}[htbp]
\figurenum{1d}
\begin{center}
\includegraphics{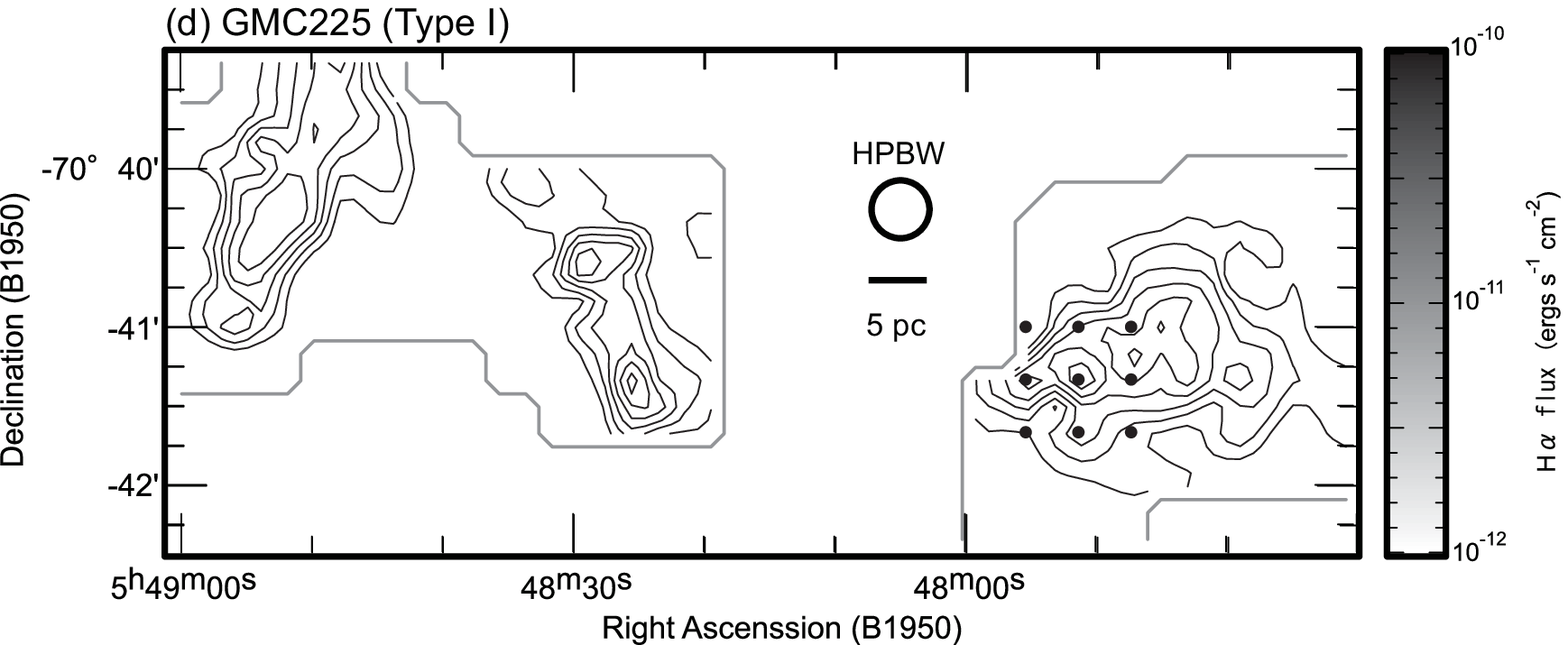}
\end{center}
\caption{ \label{fig01d}}
\end{figure}
\newpage



\begin{figure}[htbp]
\figurenum{2a}
\begin{center}
\includegraphics{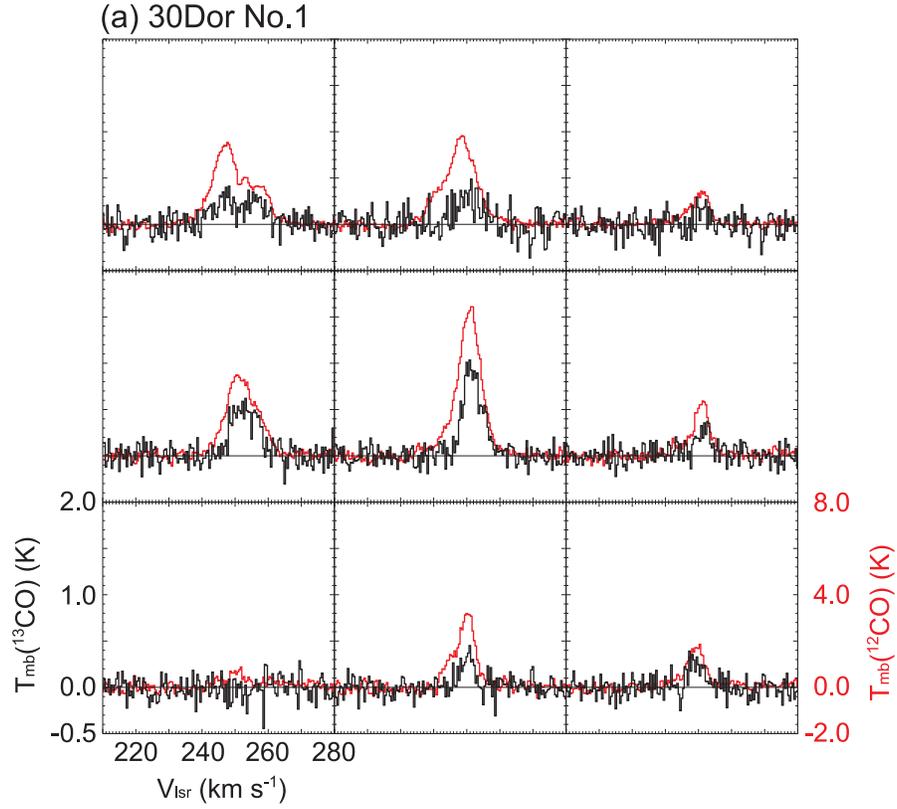}
\end{center}
\caption{
$^{13}$CO($J=3-2$) profile maps of the (a)30Dor No.1, (b)30Dor No.2, (c)30Dor No.3\&4, (d)N159 No.1, (e)N159 No.2, (f)N159 No.4, (g)N206D No.1, and (h)GMC225 No.1 overlaid with $^{12}$CO($J=3-2$) profiles. Black spectra are $^{13}$CO($J=3-2$) and red are  $^{12}$CO($J=3-2$). \label{fig02a}
}
\end{figure}
\newpage

\begin{figure}[htbp]
\figurenum{2b}
\begin{center}
\includegraphics{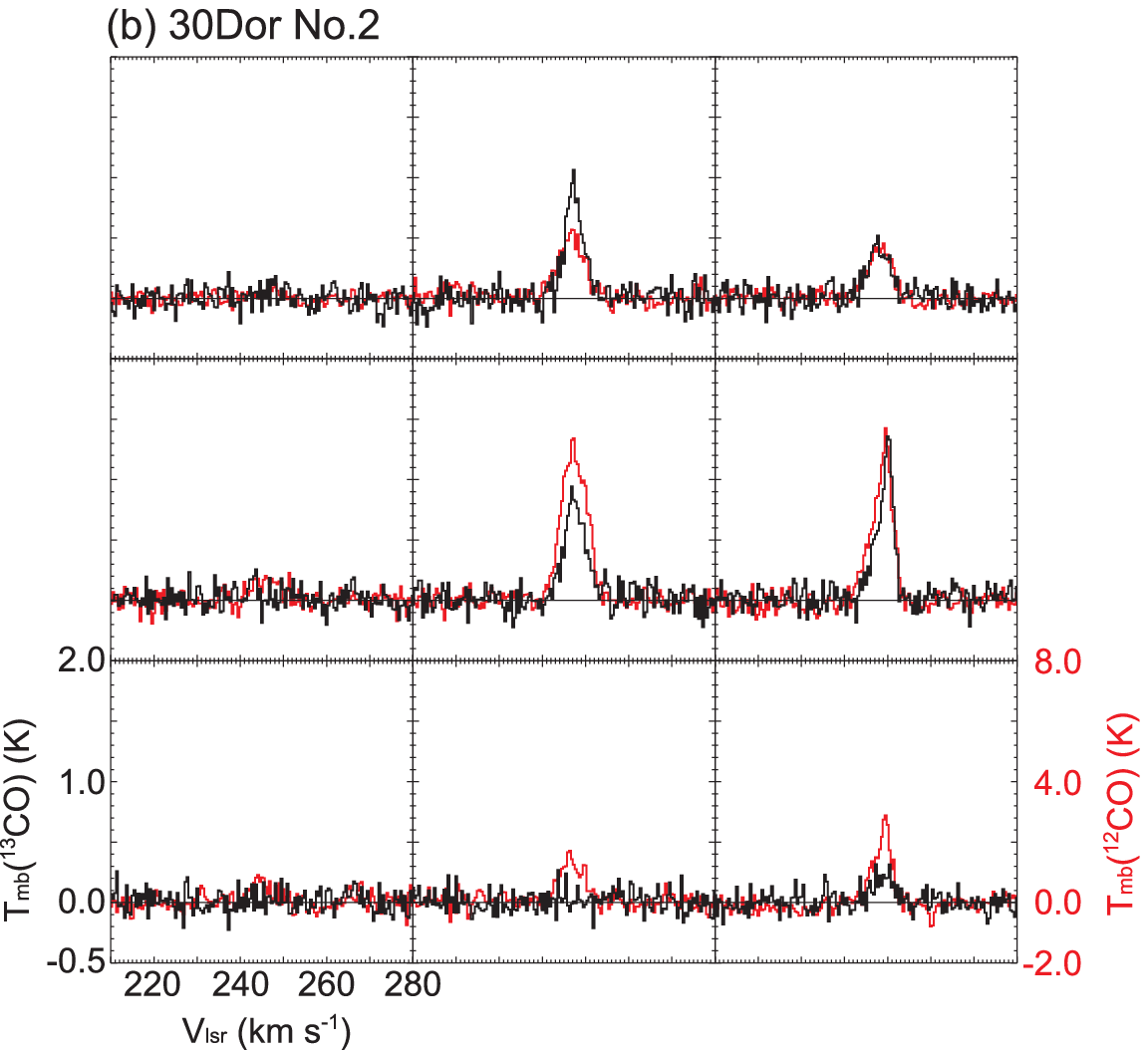}
\end{center}
\caption{ \label{fig02b}}
\end{figure}
\newpage

\begin{figure}[htbp]
\figurenum{2c}
\begin{center}
\includegraphics{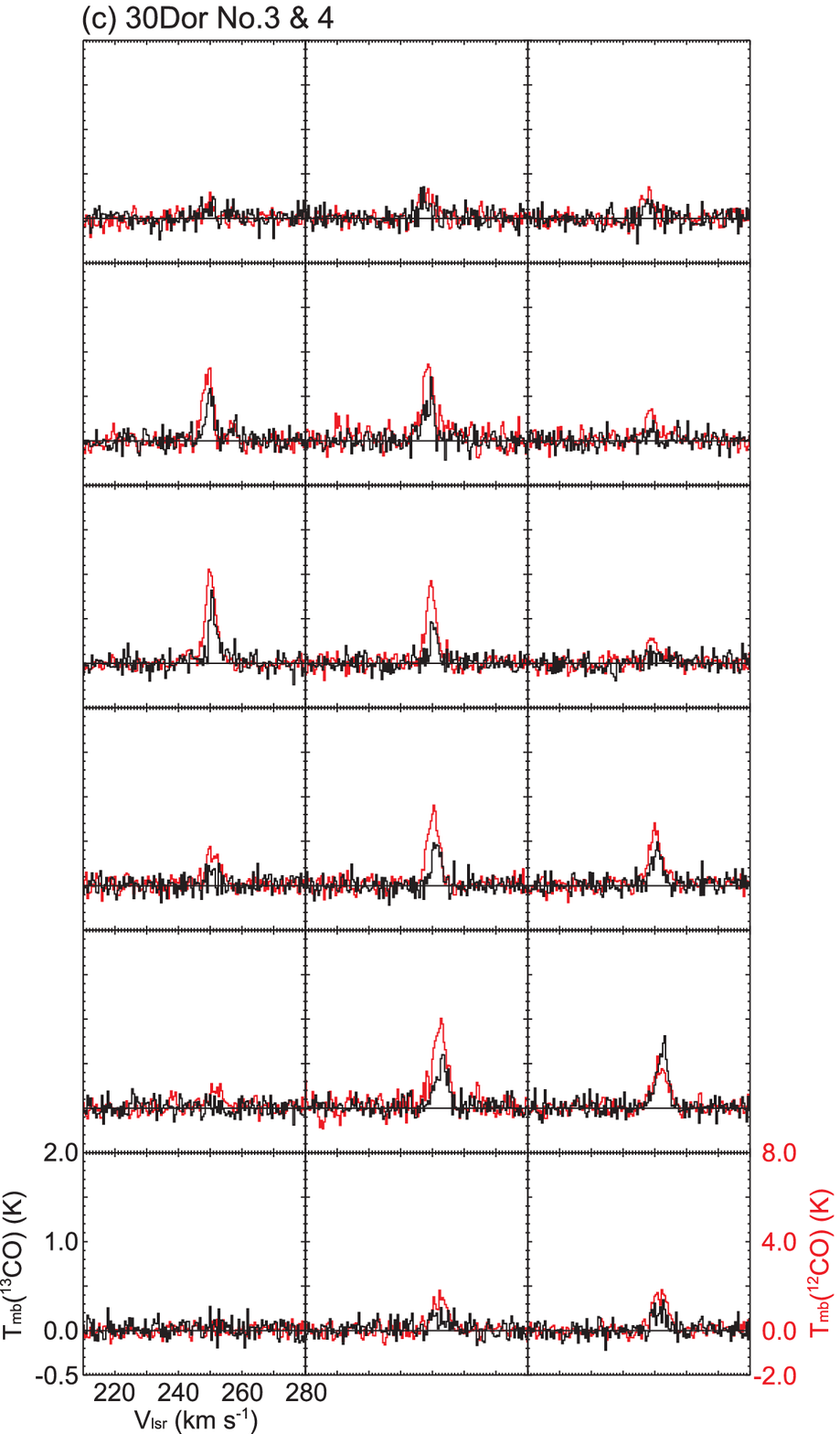}
\end{center}
\caption{ \label{fig02c}}
\end{figure}
\newpage

\begin{figure}[htbp]
\figurenum{2d}
\begin{center}
\includegraphics{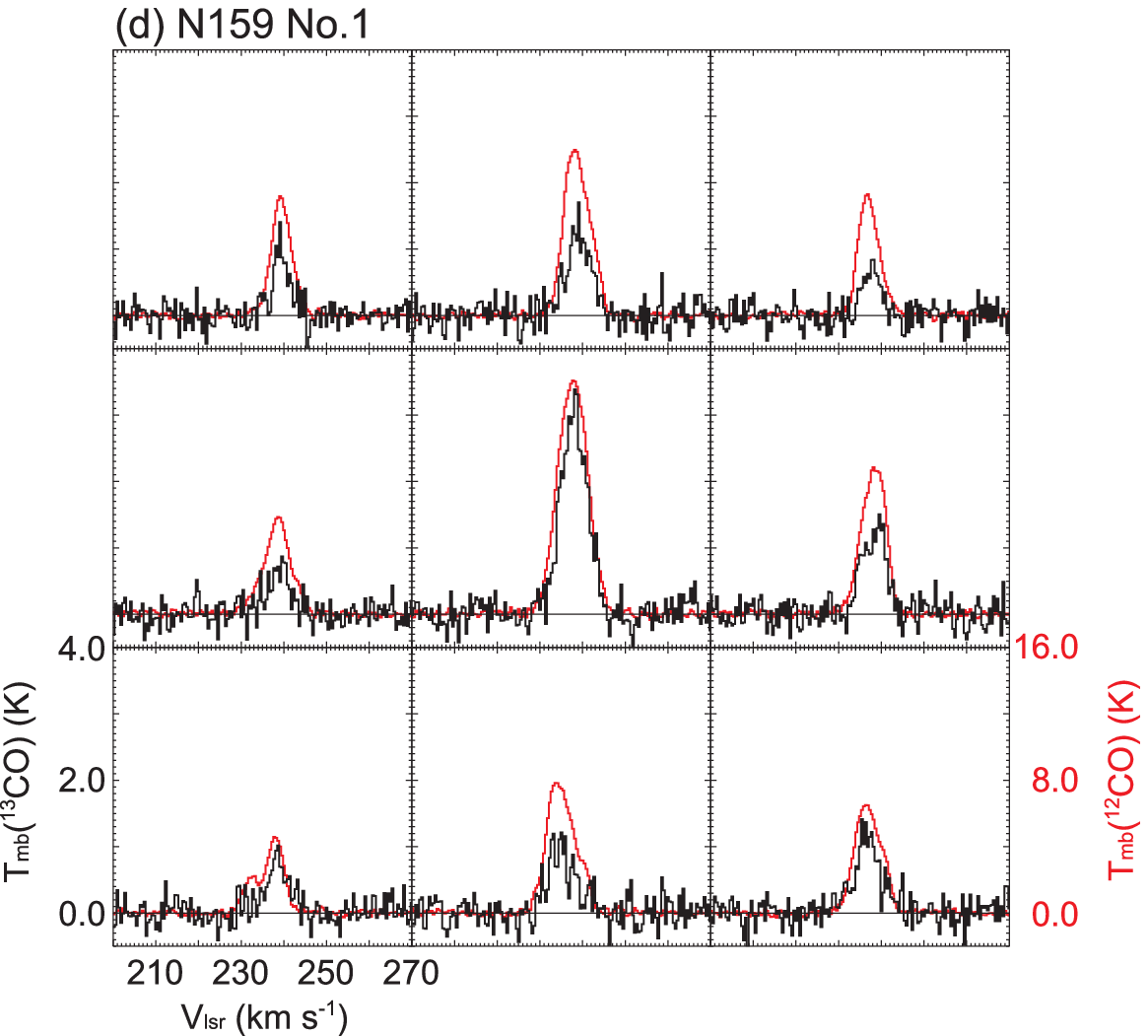}
\end{center}
\caption{ \label{fig02d}}
\end{figure}
\newpage

\begin{figure}[htbp]
\figurenum{2e}
\begin{center}
\includegraphics{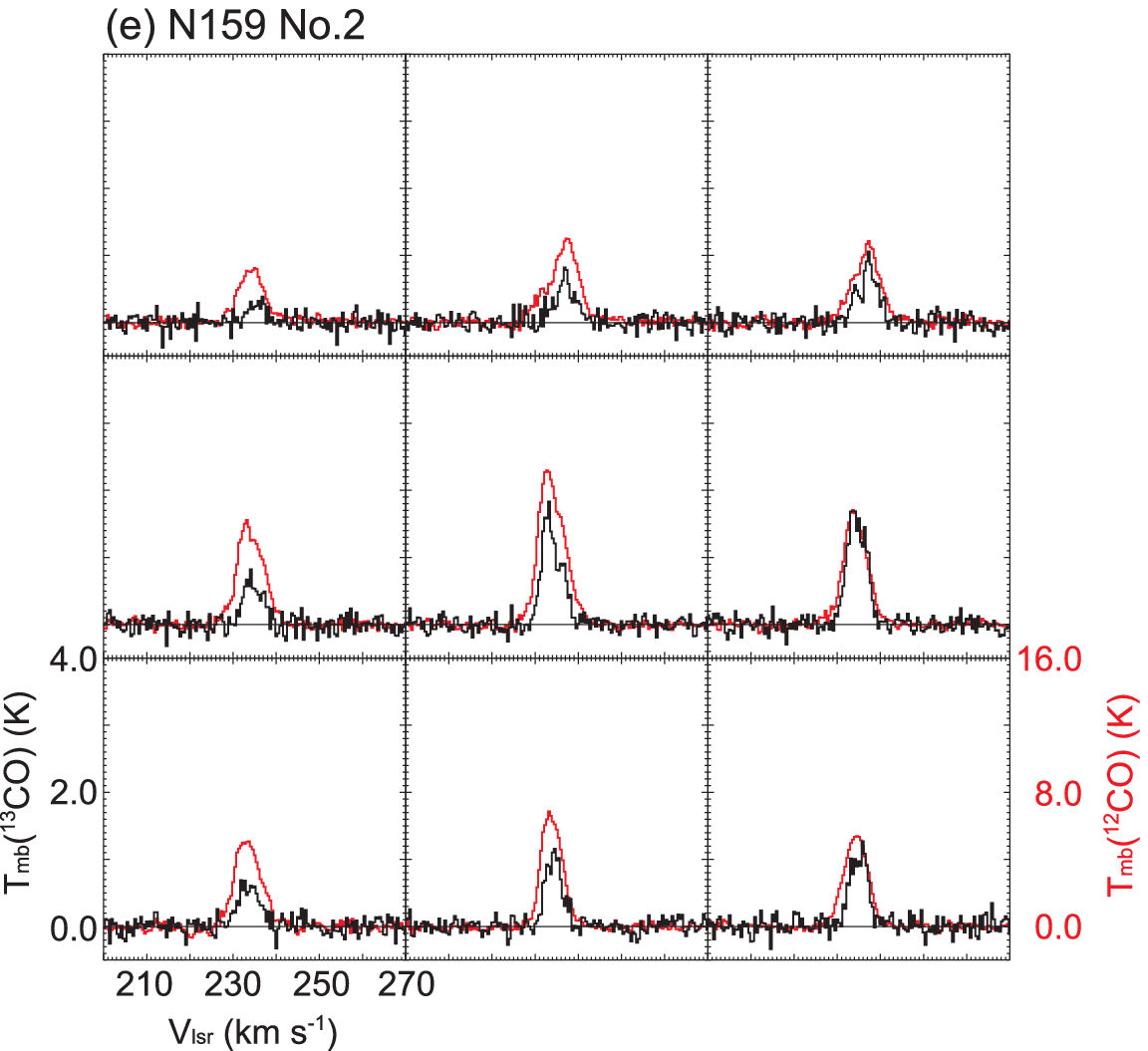}
\end{center}
\caption{ \label{fig02e}}
\end{figure}
\newpage

\begin{figure}[htbp]
\figurenum{2f}
\begin{center}
\includegraphics{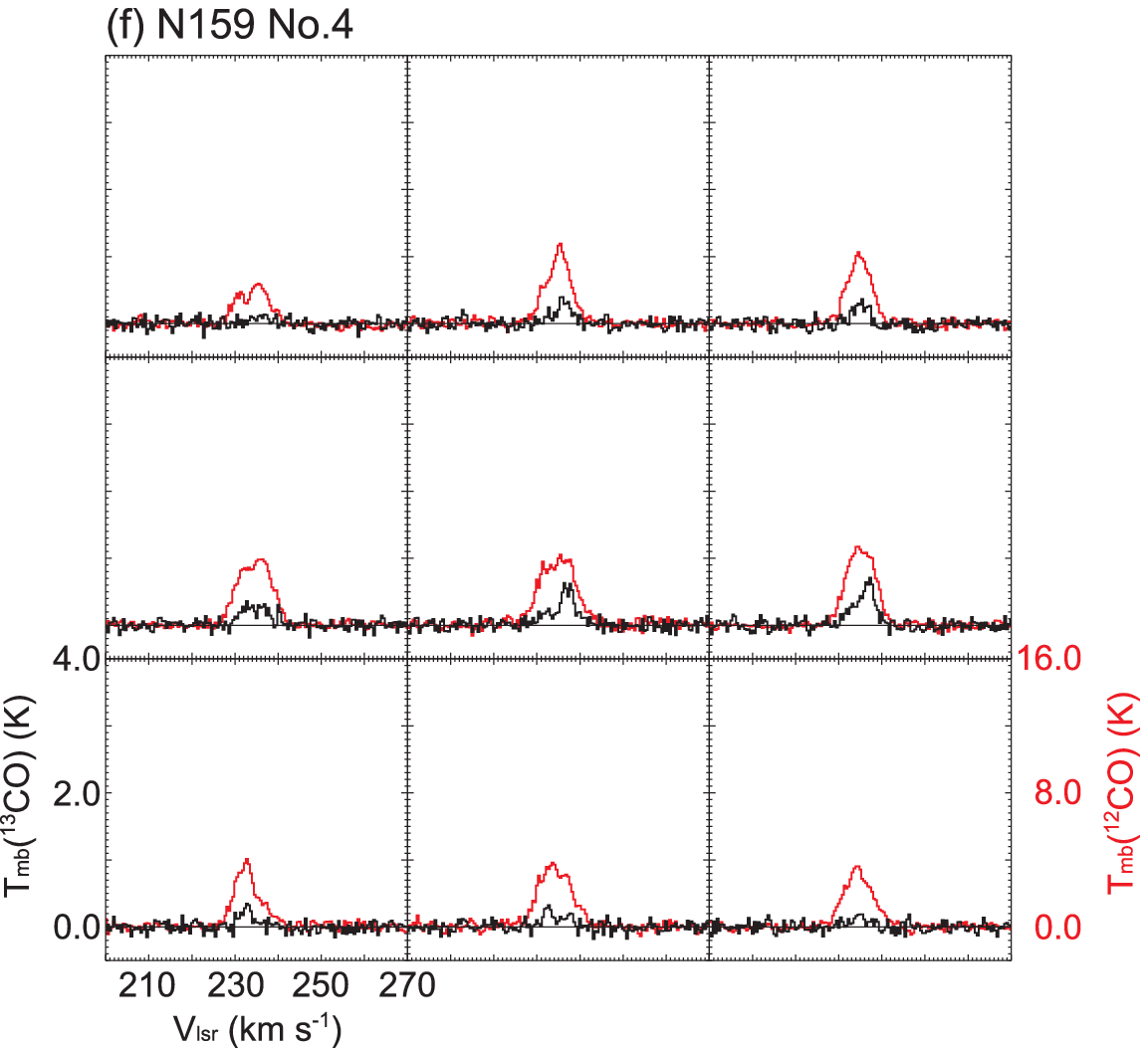}
\end{center}
\caption{ \label{fig02f}}
\end{figure}
\newpage

\begin{figure}[htbp]
\figurenum{2g}
\begin{center}
\includegraphics{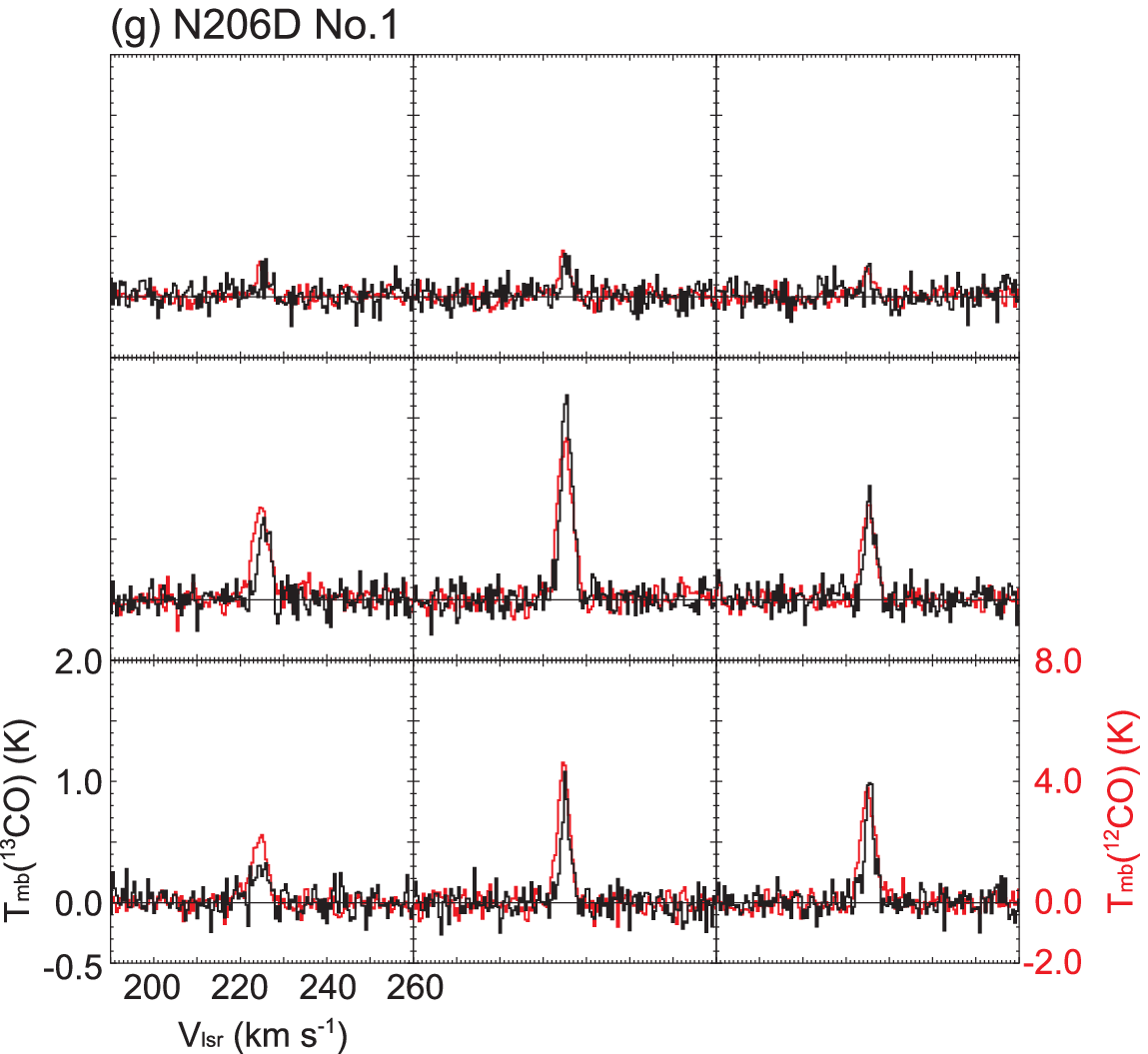}
\end{center}
\caption{ \label{fig02g}}
\end{figure}
\newpage

\begin{figure}[htbp]
\figurenum{2h}
\begin{center}
\includegraphics{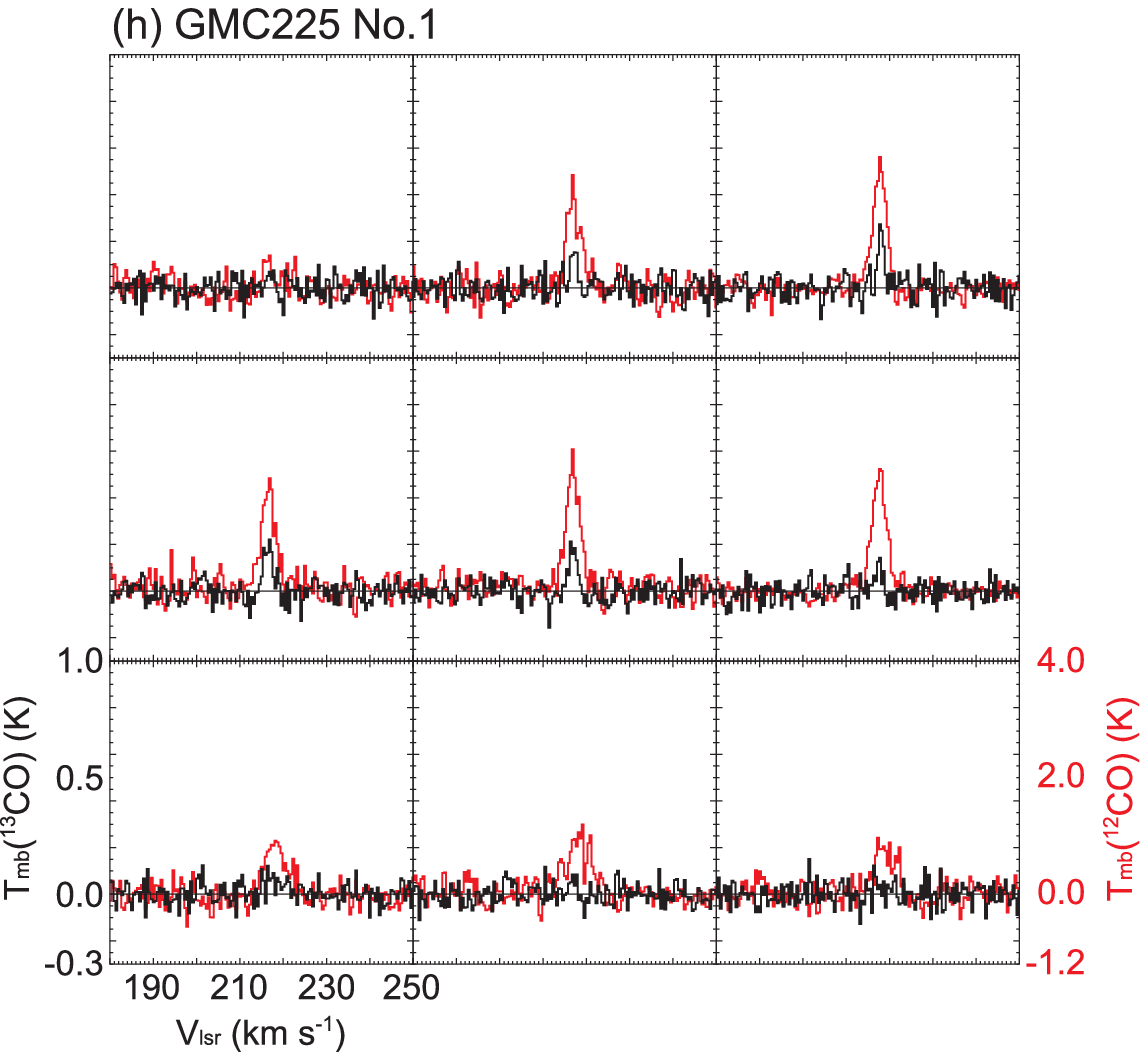}
\end{center}
\caption{ \label{fig02h}}
\end{figure}
\newpage


\setcounter{figure}{2}

\begin{figure}[htbp]
\begin{center}
\includegraphics{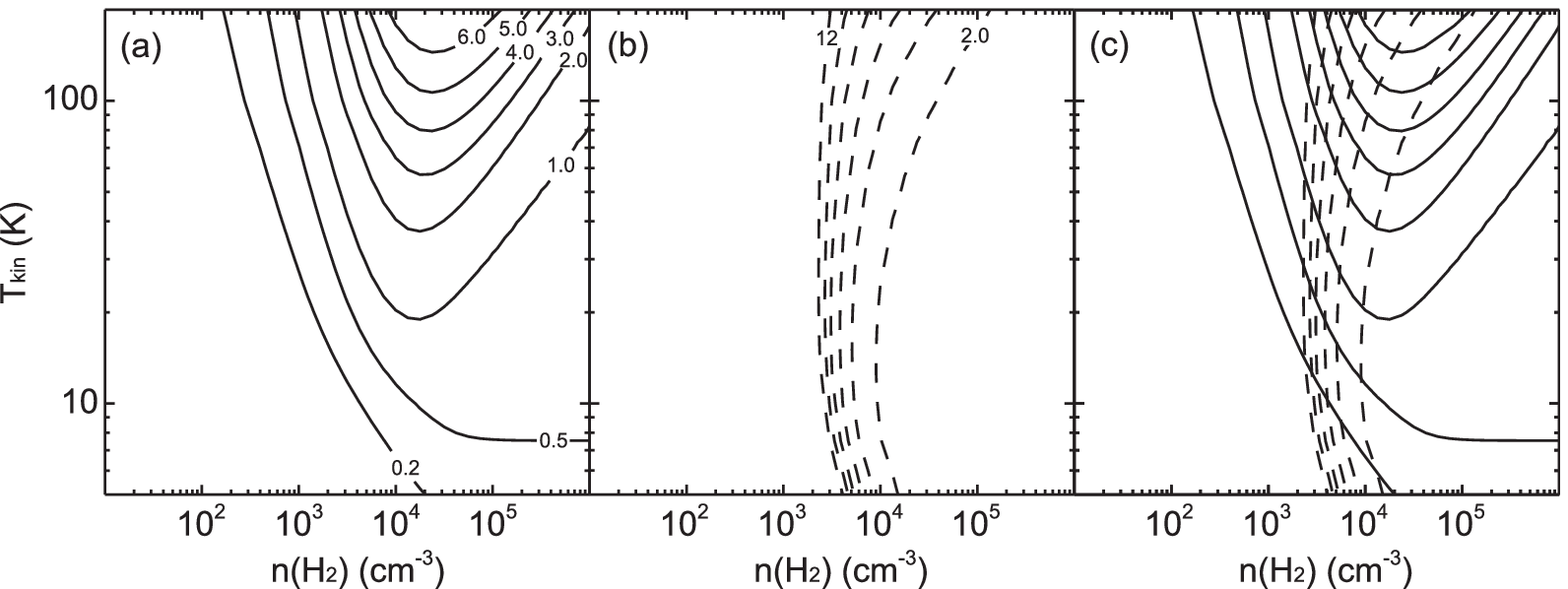}
\end{center}
\caption{
Contour plots of LVG analysis for reference. Contours are (a) $R^{13}_{3-2/1-0}$, (b) $R^{12/13}_{3-2}$, and (c) a + b. 
Here, $X$(CO)$= 1.6\times10^{-5}$, $dv/dr = 1.0$ km s$^{-1}$pc$^{-1}$, and the \tmb{abundance ratio of }{number ratio (abundance ratio)} $^{12}$CO/$^{13}$CO is 50. \label{fig03}
}
\end{figure}
\newpage

\begin{figure}[htbp]
\begin{center}
\includegraphics{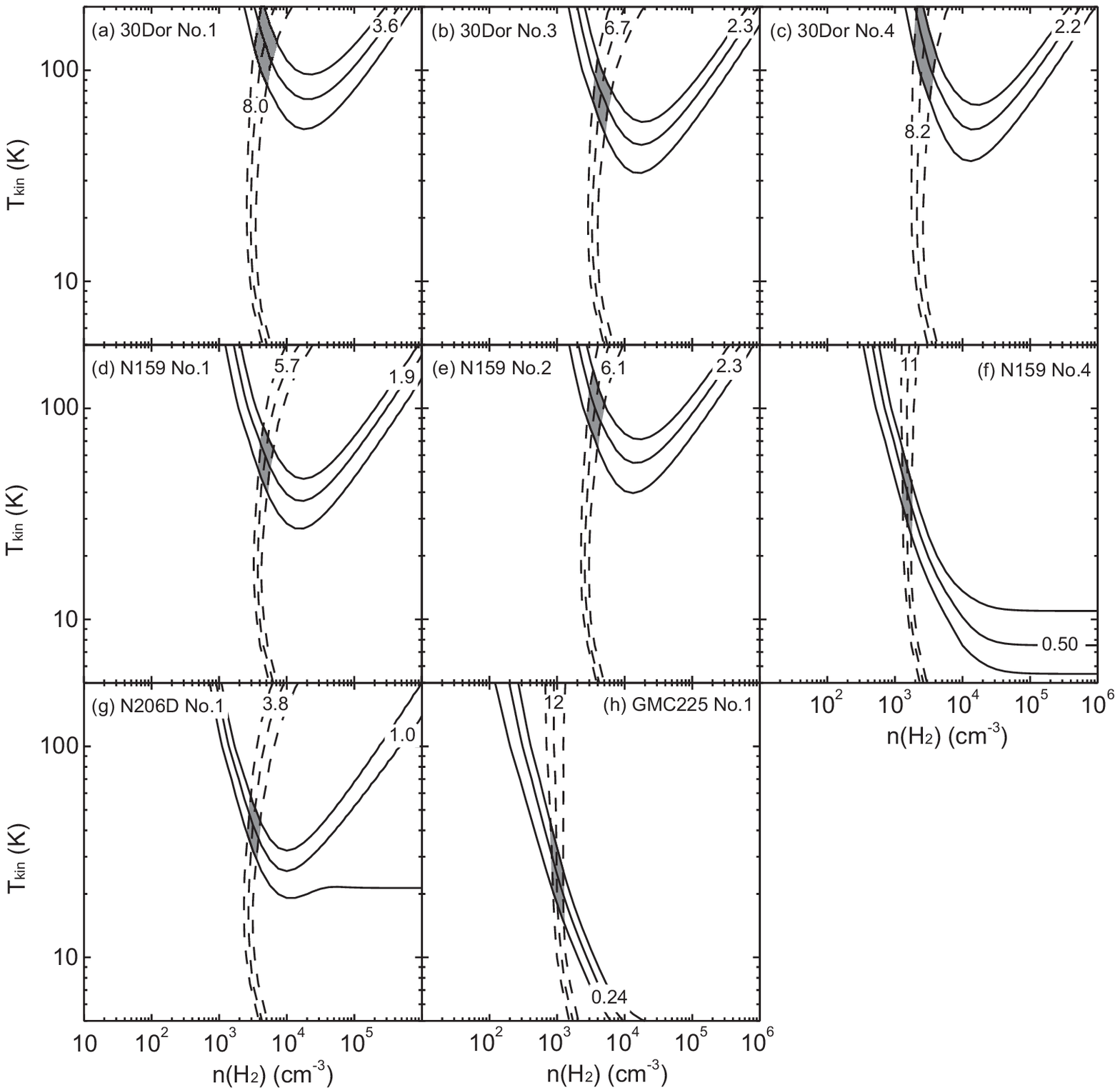}
\end{center}
\caption{
Contour plots of LVG analysis \tmb{of }{for} 8 clumps: 
(a) 30 Dor No.1, 
(b) 30 Dor No.3, 
(c) 30 Dor No.4, 
(d) N159 No.1, 
(e) N159 No.2, 
(f) N159 No.4, 
(g) N206D No.1, and 
(h) GMC225 No.1. 
The vertical axis is kinetic temperature $T_{\mathrm{kin}}$, and the horizontal axis is molecular hydrogen density $n$(H$_2$). Solid lines indicate $R^{13}_{3-2/1-0}$, and dashed lines indicate $R^{12/13}_{3-2}$. Hatched areas are the regions in which these two ratios overlap within intensity calibration errors. 
 \label{fig04}
}
\end{figure}
\newpage

\begin{figure}[htbp]
\begin{center}
\includegraphics{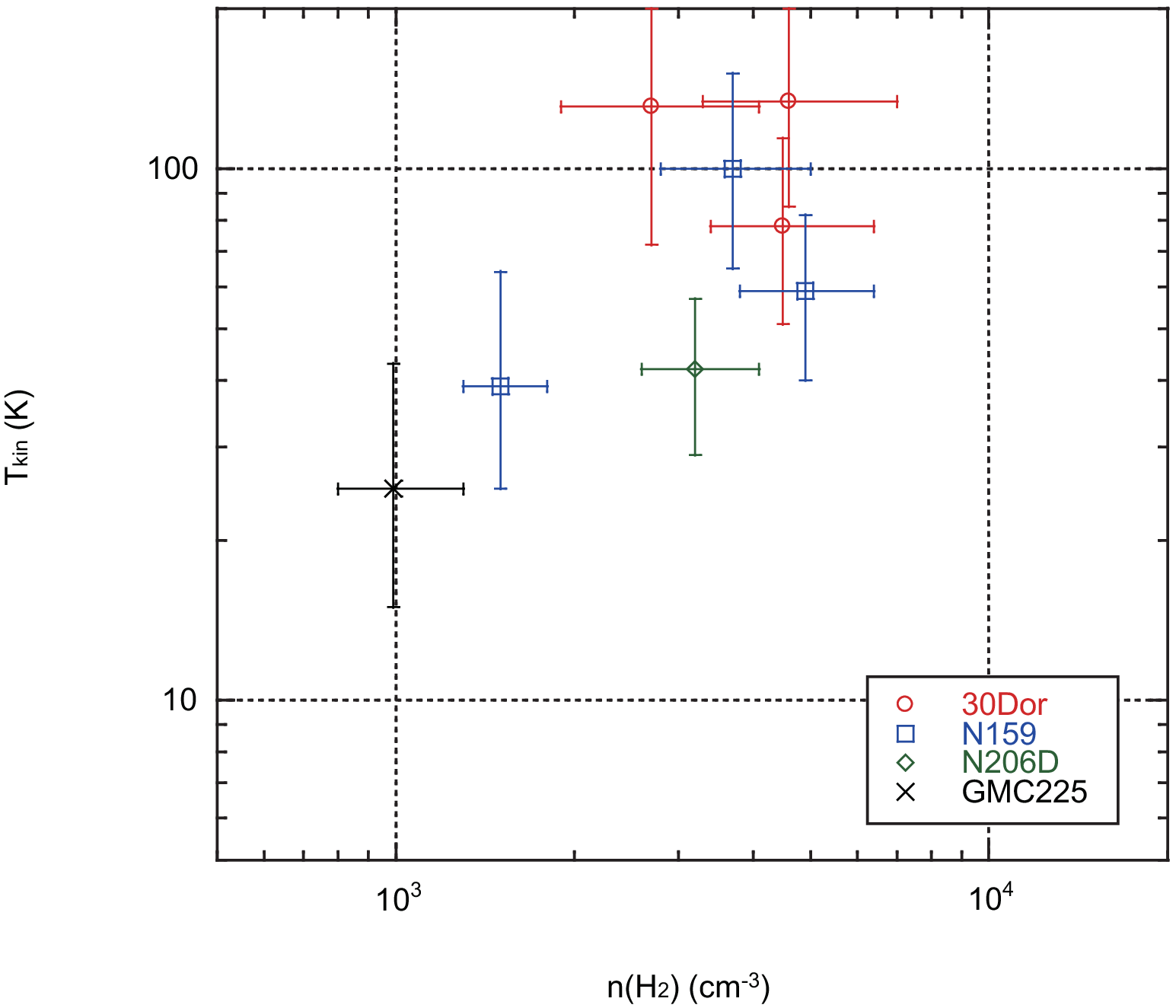}
\end{center}
\caption{
Plot of LVG results. The vertical axis is kinetic temperature, $T_{\mathrm{kin}}$, and the horizontal axis is molecular hydrogen density $n$(H$_2$).
 \label{fig05}
}
\end{figure}
\newpage

\begin{figure}[htbp]
\begin{center}
\includegraphics{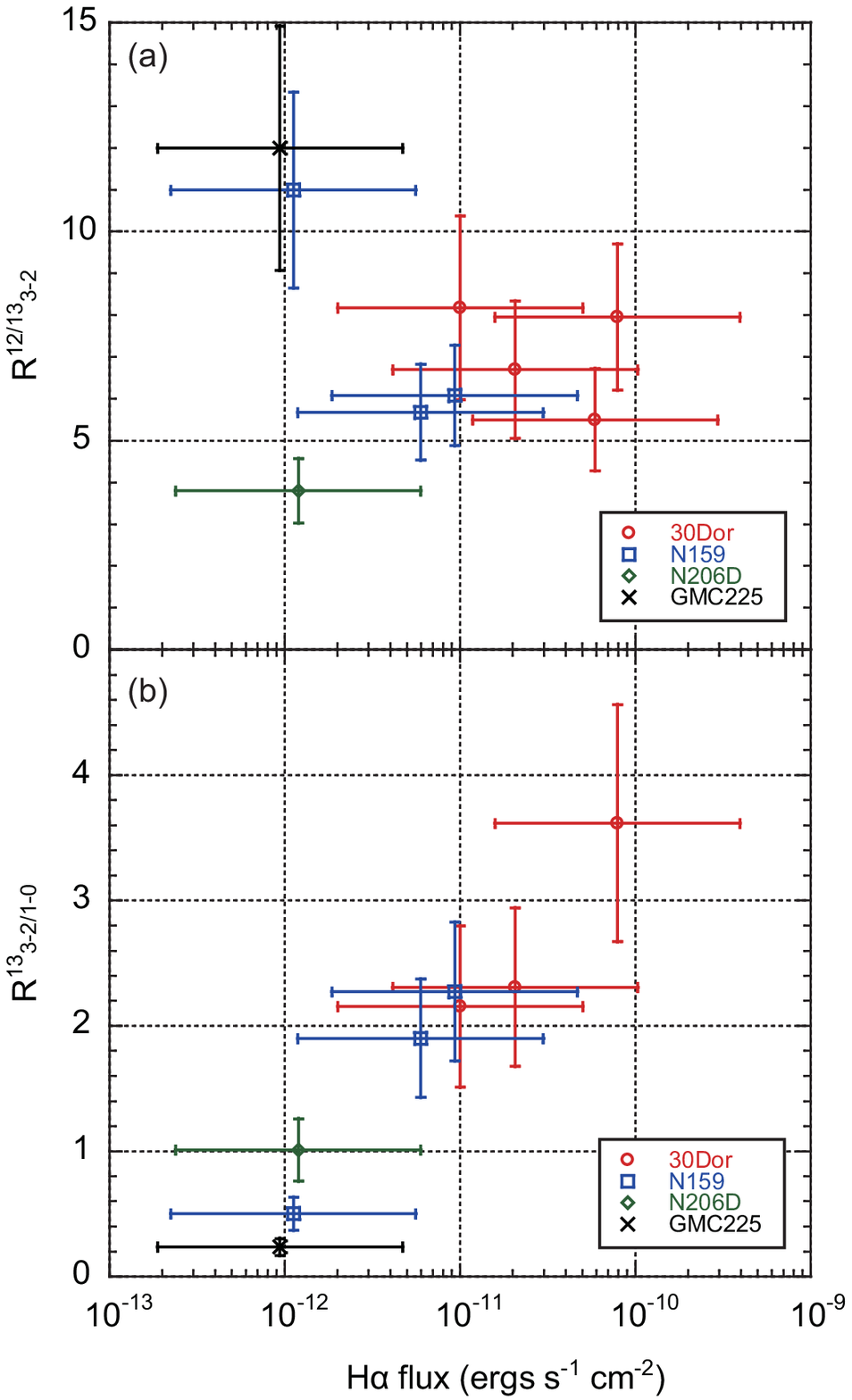}
\end{center}
\caption{
Relations between line intensity ratio and H$\alpha$ flux. (a) Intensity ratio of $^{12}$CO($J=3-2$) to $^{13}$CO($J=3-2$), $R^{12/13}_{3-2}$. (b) Intensity ratio of $^{13}$CO($J=3-2$) to $^{13}$CO($J=1-0$), $R^{13}_{3-2/1-0}$. \label{fig06}
}
\end{figure}
\newpage

\begin{figure}[htbp]
\begin{center}
\includegraphics{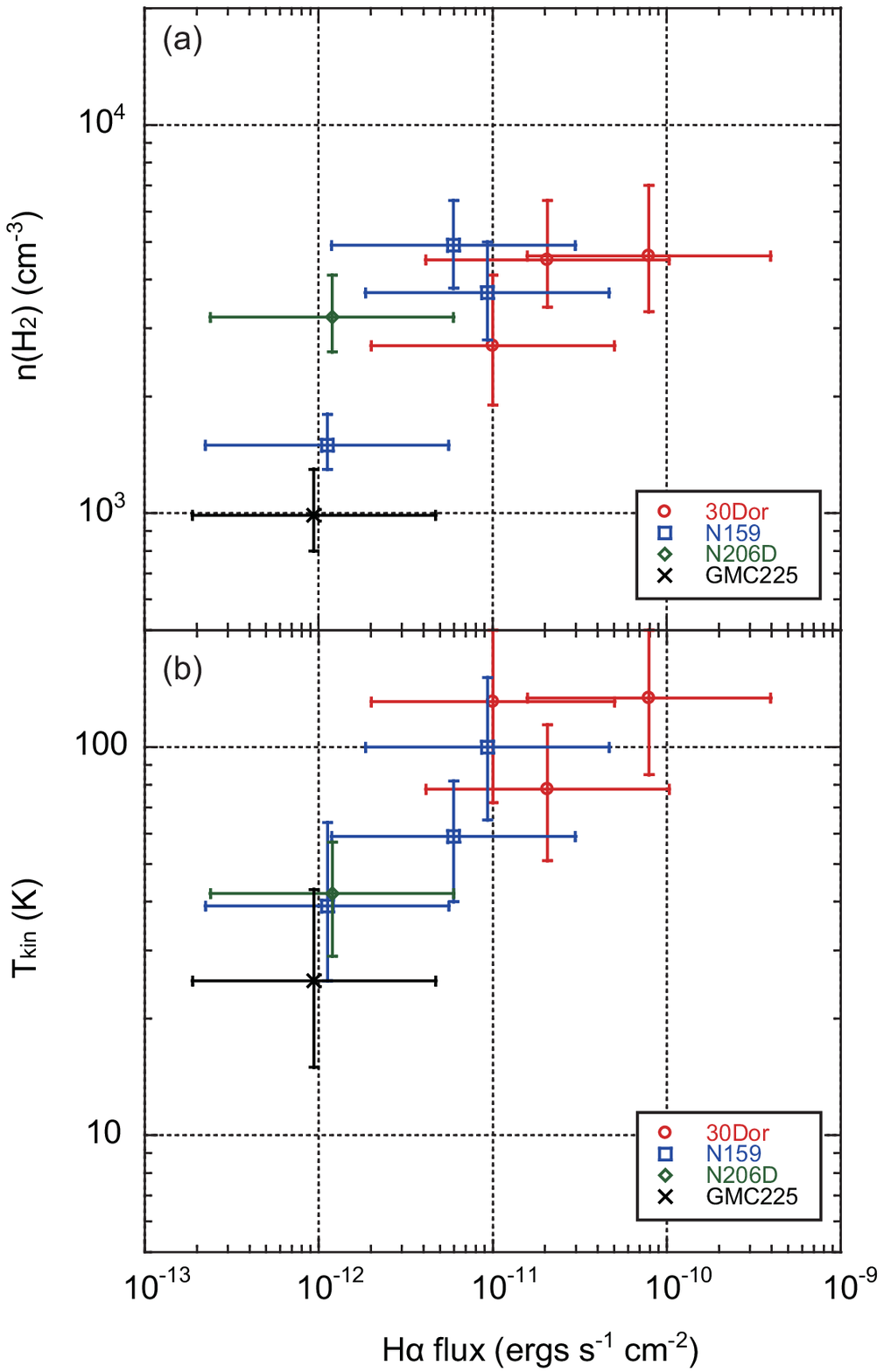}
\end{center}
\caption{
Plot of physical properties as a function of H$\alpha$ flux at $^{12}$CO($J=3-2$) peak. The vertical axises are (a) molecular hydrogen density $n$(H$_2$), and (b) kinetic temperature, $T_{\mathrm{kin}}$. The horizontal axis is H$\alpha$ flux.
 \label{fig07}
}
\end{figure}
\newpage

\clearpage

\begin{deluxetable}{ccccccccccccccc}
 \tabletypesize{\scriptsize}
 \rotate
 \tablewidth{0pc}
 \tablecaption{Observerd Clumps and Properties of $^{13}$CO($J=3-2$) line. \label{tab1}}

\tablehead{
  \multicolumn{3}{c}{GMC} &

  \colhead{} & 
  \multicolumn{4}{c}{Clump\tablenotemark{c}} &

  \colhead{} & 
  \multicolumn{5}{c}{Peak Properties of $^{13}$CO($J=3-2$) Line} &

  \colhead{} \\
  
  \cline{1-3} \cline{5-8} \cline{10-14}
  
  \colhead{No.\tablenotemark{a}} & 
  \colhead{Name\tablenotemark{a}} & 
  \colhead{Type\tablenotemark{b}} & 
  \colhead{} &
  \colhead{} & 
  \colhead{} & 
  \multicolumn{2}{c}{Position} & 
  
  \colhead{} & 
  \colhead{$T_{\mathrm{mb}}$} & 
  \colhead{$V_{\mathrm{lsr}}$} & 
  \colhead{$\Delta V$} & 
  \colhead{Integrated Intensity} & 
  \colhead{rms} & 
  \colhead{Other ID\tablenotemark{d}} \\
  
  \cline{7-8}
  
  \colhead{} & 
  \colhead{} & 
  \colhead{} & 
  \colhead{} & 
  \colhead{Region} & 
  \colhead{No.} & 
  \colhead{$\alpha$(1950)} & 
  \colhead{$\delta$(1950)} & 
  \colhead{} & 
  \colhead{(K)} & 
  \colhead{(km s$^{-1}$)} & 
  \colhead{(km s$^{-1}$)} & 
  \colhead{(K km s$^{-1}$)} & 
  \colhead{(K)} & 
  \colhead{} \\
  
  \colhead{(1)} & 
  \colhead{(2)} & 
  \colhead{(3)} & 
  \colhead{} &
  \colhead{(4)} & 
  \colhead{(5)} & 
  \colhead{(6)} & 
  \colhead{(7)} & 
  \colhead{} & 
  \colhead{(8)} & 
  \colhead{(9)} & 
  \colhead{(10)} & 
  \colhead{(11)} & 
  \colhead{(12)} & 
  \colhead{(13)} 
}

\startdata
 186 & LMC N J0538-6904 & III & & 30 Dor & 1 & 5 39 08.6 & -69 06 15 & & 0.99 & 251.3 & 5.5 &  7.1  & 0.10 & 30Dor-10 \\
     &                  &     & &        & 2 & 5 38 54.6 & -69 08 00 & & 0.86 & 247.4 & 5.2 &  5.2  & 0.09 & 30Dor-12 \\
     &                  &     & &        & 3 & 5 38 49.0 & -69 04 30 & & 0.54 & 253.3 & 3.8 &  2.9  & 0.08 & 30Dor-06 \\
     &                  &     & &        & 4 & 5 38 49.0 & -69 03 30 & & 0.47 & 249.0 & 3.8 &  2.2  & 0.08 &          \\
 197 & LMC N J0540-7008 & III & & N159   & 1 & 5 40 03.7 & -69 47 00 & & 3.2  & 237.9 & 5.8 & 23    & 0.18 & N159W    \\
     &                  &     & &        & 2 & 5 40 35.5 & -69 46 00 & & 1.7  & 233.1 & 3.7 &  8.4  & 0.11 & N159E    \\
     &                  &     & &        & 4 & 5 40 32.7 & -69 52 00 & & 0.60 & 237.0 & 3.5 &  3.2  & 0.07 & N159S    \\
 156 & LMC N J0532-7114 & II  & & N206D  & 1 & 5 32 58.4 & -71 15 20 & & 1.6  & 225.4 & 3.0 &  5.7  & 0.09 &          \\
 225 & LMC N J0547-7041 & I   & & GMC225 & 1 & 5 47 51.3 & -70 41 20 & & 0.19 & 216.9 & 2.2 &  0.33 & 0.04 &          \\
\enddata

\tablecomments{
Units of right ascension are hours, minutes, seconds, and units of declination are degrees, arcminites, arcseconds. 
Col. (1): Running number of GMC used in Table 1 in Fukui et al. (2008). 
Col. (2): Name of GMC.
Col. (3): Type of GMC.
Col. (4): Region name used in this paper.
Col. (5): Running number of $^{12}$CO($J=3-2$) clump in each region used in Table 2 in Minamidani et al. (2008). 
Cols. (6)--(7): Coordinates of the position of $^{12}$CO($J=3-2$) clump. 
Cols. (8)--(12): Observed properties of the $^{13}$CO($J=3-2$) spectra obtained at the peak positions of the $^{12}$CO($J=3-2$) clumps. 
The peak main-beam temperature $T_{\mathrm{mb}}$, $V_{\mathrm{LSR}}$, and the FWHM line width $\Delta V$ are 
derived from a single Gaussian curve fitting and are given in cols. (8), (9), and (10), respectively. 
The $^{13}$CO($J=3-2$) integrated inteisities and r.m.s. noise level at the peak positions of the $^{12}$CO($J=3-2$) clumps are shown in cols. (11) and (12), respectively. 
Col. (13): Another identification based on $^{12}$CO($J=1-0$) observations with SEST.
}
\tablenotetext{a}{Fukui et al. (2008).}
\tablenotetext{b}{Kawamura et al. (2009).}
\tablenotetext{c}{Minamidani et al. (2008).}
\tablenotetext{d}{Johansson et al. (1998).}

\end{deluxetable}

\clearpage

\begin{deluxetable}{ccccccccccc}
 \tabletypesize{\scriptsize}
 \rotate
 \tablewidth{0pc}
 \tablecaption{Line Intensities and Line Ratios. \label{tab2}}

\tablehead{
 \colhead{} & 
 \colhead{} & 
 \multicolumn{2}{c}{Convolved $T_{\mathrm{mb}}$ (K)} & 
 
 \colhead{} & 
 \multicolumn{2}{c}{$T_{\mathrm{mb}}$ (K)} & 
 
 \colhead{} & 
 \multicolumn{2}{c}{Line Ratio} & 
 
 \colhead{H$\alpha$ flux} \\
 
  \cline{3-4} \cline{6-7} \cline{9-10}

 \colhead{Region} & 
 \colhead{No.} & 
 \colhead{$^{13}$CO($J=3-2$)} & 
 \colhead{$^{12}$CO($J=3-2$)} & 
 \colhead{} & 
 \colhead{$^{13}$CO($J=1-0$)} & 
 \colhead{Ref.} & 
 \colhead{} & 
 \colhead{$R^{12/13}_{3-2}$} & 
 \colhead{$R^{13}_{3-2/1-0}$} & 
 \colhead{($\times 10^{-12}$ ergs s$^{^1}$ cm$^{-2}$)} \\
 
 \colhead{(1)} & 
 \colhead{(2)} & 
 \colhead{(3)} & 
 \colhead{(4)} & 
 \colhead{} & 
 \colhead{(5)} & 
 \colhead{(6)} & 
 \colhead{} & 
 \colhead{(7)} & 
 \colhead{(8)} & 
 \colhead{(9)} 
}

\startdata
 30 Dor & 1 & 0.47$\pm$0.08        & 3.7$\pm$0.5 & & 0.13 & J98 & & \phn8.0$\pm$1.7          & 3.6\phn$\pm$0.9\phn & 79\phantom{.00} \\
        & 2 & 0.46$\pm$0.08        & 2.5$\pm$0.4 & & -    & J98 & & \phn5.5$\pm$1.2          & -                   & 59\phantom{.00} \\
        & 3 & 0.30$\pm$0.06        & 2.0$\pm$0.3 & & 0.13 & J98 & & \phn6.7$\pm$1.6          & 2.3\phn$\pm$0.6\phn & 21\phantom{.00} \\
        & 4 & 0.28$\pm$0.06        & 2.3$\pm$0.4 & & 0.13 & J98 & & \phn8.2$\pm$2.2          & 2.2\phn$\pm$0.6\phn & 10\phantom{.00} \\
 N159   & 1 & 1.5\phn$\pm$0.2\phn & 8.6$\pm$1.2 & & 0.80 & J98 & & \phn5.7$\pm$1.1          & 1.9\phn$\pm$0.5\phn & \phantom{0}6.0\phantom{0} \\
        & 2 & 1.0\phn$\pm$0.1\phn & 6.1$\pm$0.8 & & 0.44 & J98 & & \phn6.1$\pm$1.2          & 2.3\phn$\pm$0.6\phn & \phantom{0}9.3\phantom{0} \\
        & 4 & 0.36$\pm$0.06        & 3.8$\pm$0.5 & & 0.72 & J98 & & 11\phd\phn$\pm$2\phd\phn & 0.50$\pm$0.13       & \phantom{0}1.1\phantom{0} \\
 N206D  & 1 & 0.86$\pm$0.12        & 3.3$\pm$0.5 & & 0.85 & M08 & & \phn3.8$\pm$0.8          & 1.0\phn$\pm$0.2\phn & \phantom{0}1.2\phantom{0} \\
 GMC225 & 1 & 0.13$\pm$0.03        & 1.5$\pm$0.2 & & 0.55 & M08 & & 12\phd\phn$\pm$3\phd\phn & 0.24$\pm$0.07       & \phantom{0}0.94 \\

\enddata

\tablecomments{
Col. (1): Region. 
Col. (2): Running number in each region. 
Cols. (3)-(4): The peak main-beam temperature, $T_{\mathrm{mb}}$, of the $^{13}$CO($J=3-2$) and $^{12}$CO($J=3-2$), respectively, derived by using a single Gaussian curve fitting for a spectrum obtaind by convolved spectra into the 45\arcsec beam with a Gaussian kernel.
Cols. (5)-(6): The peak main-beam temperature, $T_{\mathrm{mb}}$, of the $^{13}$CO($J=1-0$) (col. (5)) and their references (col. (6)).
Cols. (7)-(8): Ratios of the peak main-beam temperatures. Ratios of $^{12}$CO($J=3-2$) to $^{13}$CO($J=3-2$) are shown in col. (7), and ratios of $^{13}$CO($J=3-2$) to $^{13}$CO($J=1-0$) are shown in col. (8).
Col. (9): H$\alpha$ flux toward the peak positions. 
}
\tablerefs{(J98) Johansson et al. 1998; (M08) Minamidani et al. 2008.}

\end{deluxetable}

\clearpage

\begin{deluxetable}{cccccccc}
 \tabletypesize{\footnotesize}
 \rotate
 \tablewidth{0pc}
 \tablecaption{LVG results. \label{tab3}}

\tablehead{
 \colhead{Region} & 
 \colhead{No.} & 
 \colhead{$dv/dr$} & 
 \colhead{$R^{12/13}_{3-2}$} & 
 \colhead{$R^{13}_{3-2/1-0}$} & 
 \colhead{$n$(H$_2$)} & 
 \colhead{$T_{\mathrm{kin}}$} & 
 \colhead{H$\alpha$ flux} \\ 
  
 \colhead{} & 
 \colhead{} & 
 \colhead{(km s$^{-1}$ pc$^{-1}$)} & 
 \colhead{} & 
 \colhead{} & 
 \colhead{($\times 10^3$ cm$^{-3}$)} & 
 \colhead{(K)} & 
 \colhead{($\times 10^{-12}$ ergs s$^{-1}$ cm$^{-2}$)} \\
  
 \colhead{(1)} & 
 \colhead{(2)} & 
 \colhead{(3)} & 
 \colhead{(4)} & 
 \colhead{(5)} & 
 \colhead{(6)} & 
 \colhead{(7)} & 
 \colhead{(8)}
}

\startdata

 30 Dor & 1 & 0.9 & \phn8.0$\pm$1.7      & 3.6\phn$\pm$0.9\phn & $4.6\phn^{+2.4\phn}_{-1.3\phn}$ & $134^{+66}_{-49}$    & 79\phantom{.00} \\
        & 3 & 0.9 & \phn6.7$\pm$1.6      & 2.3\phn$\pm$0.6\phn & $4.5\phn^{+1.9\phn}_{-1.1\phn}$ & $\phn78^{+36}_{-27}$ & 21\phantom{.00} \\
        & 4 & 0.5 & \phn8.2$\pm$2.2      & 2.2\phn$\pm$0.6\phn & $2.7\phn^{+1.4\phn}_{-0.8\phn}$ & $131^{+69}_{-59}$    & 10\phantom{.00} \\
 N159   & 1 & 0.9 & \phn5.7$\pm$1.1      & 1.9\phn$\pm$0.5\phn & $4.9\phn^{+1.5\phn}_{-1.1\phn}$ & $\phn59^{+23}_{-19}$ & \phantom{0}6.0\phantom{0} \\
        & 2 & 0.5 & \phn6.1$\pm$1.2      & 2.3\phn$\pm$0.6\phn & $3.7\phn^{+1.3\phn}_{-0.9\phn}$ & $100^{+51}_{-35}$    & \phantom{0}9.3\phantom{0} \\
        & 4 & 0.4 & 11\phd\phn$\pm$2\phn & 0.50$\pm$0.13       & $1.5\phn^{+0.3\phn}_{-0.2\phn}$ & $\phn39^{+25}_{-14}$ & \phantom{0}1.1\phantom{0} \\
 N206D  & 1 & 0.3 & \phn3.8$\pm$0.8      & 1.0\phn$\pm$0.2\phn & $3.2\phn^{+0.9\phn}_{-0.6\phn}$ & $\phn42^{+15}_{-13}$ & \phantom{0}1.2\phantom{0} \\
 GMC225 & 1 & 0.2 & 12\phd\phn$\pm$3\phn & 0.24$\pm$0.07       & $0.99^{+0.31}_{-0.19}$          & $\phn25^{+18}_{-10}$ & \phantom{0}0.94 \\
 
\enddata

\tablecomments{
Col. (3): Velocity gradient of clumps (Minamidani et al. 2008). 
Cols. (6) and (7): Results of LVG analysis. 
}

\end{deluxetable}

\clearpage

\begin{deluxetable}{cccccccccc}
 \rotate
 \tablewidth{0pc}
 \tablecaption{Summary of an excitation analysis in the N159 region. \label{tab4}}

\tablehead{
  \colhead{} & 
  \colhead{} & 
  \multicolumn{2}{c}{No.1 (N159W)} & 
  
  \colhead{} & 
  \multicolumn{2}{c}{No.2 (N159E)} & 
  
  \colhead{} & 
  \multicolumn{2}{c}{No.4 (N159S)} \\

    \cline{3-4} \cline{6-7} \cline{9-10}

  \colhead{Paper} & 
  \colhead{Reference} & 
  \colhead{$T_{\mathrm{kin}}$} & 
  \colhead{$n$(H$_2$)} & 
  \colhead{} & 
  \colhead{$T_{\mathrm{kin}}$} & 
  \colhead{$n$(H$_2$)} & 
  \colhead{} & 
  \colhead{$T_{\mathrm{kin}}$} & 
  \colhead{$n$(H$_2$)} \\

  \colhead{} & 
  \colhead{} & 
  \colhead{(K)} & 
  \colhead{(10$^3$ cm$^{-3}$)} & 
  \colhead{} & 
  \colhead{(K)} & 
  \colhead{(10$^3$ cm$^{-3}$)} &
  \colhead{} & 
  \colhead{(K)} & 
  \colhead{(10$^3$ cm$^{-3}$)} \\

  \colhead{(1)} & 
  \colhead{(2)} & 
  \colhead{(3)} & 
  \colhead{(4)} & 
  \colhead{} & 
  \colhead{(5)} & 
  \colhead{(6)} & 
  \colhead{} & 
  \colhead{(7)} & 
  \colhead{(8)}
}

\startdata
  This work    & --  & $59^{+23}_{-19}$ & $4.9^{+1.5}_{-1.1}$ & & $100^{+51}_{-35}$ & $3.7^{+1.3}_{-0.9}$ & & $39^{+25}_{-14}$ & $1.5^{+0.3}_{-0.2}$ \\
  Mizuno10     & M10 & $72^{+9}_{-9}$   & $4.0^{+0}_{-0.8}$   & & $83^{+26}_{-20}$  & $3.1^{+1.9}_{-0.6}$ & & $31^{+10}_{-8}$  & $1.6^{+0.4}_{-0.3}$ \\
  Pineda08     & P08 & $80$             & $10 - 100$          & & --                & --                  & & --               & --                  \\
  Minamidani08 & M08 & $> 30$           & $3 - 800$           & & $> 40$            & $1 - 300$           & & $20 - 60$        & $1 - 6$             \\
\enddata

\tablerefs{(M10) Mizuno et al. 2010; (P08) Pineda et al. 2008; (M08) Minamidani et al. 2008}

\end{deluxetable}




\end{document}